\documentclass[english,preprint]{elsarticle}
\usepackage[T1]{fontenc}
\usepackage[latin9]{inputenc}

\usepackage{textcomp}
\usepackage{pmboxdraw}
\usepackage{amsmath}
\usepackage{amssymb}
\usepackage{graphicx}
\usepackage{babel}
\begin{document}

\begin{frontmatter}{}

\title{Weak values in collision theory}

\author{Leonardo Andreta de Castro\corref{cor1}}

\ead{leonardo.castro@usp.br}

\author{Carlos Alexandre Brasil\corref{cor0}}

\author{Reginaldo de Jesus Napolitano\corref{cor0}}

\address{São Carlos Institute of Physics, University of São Paulo, PO Box
369, 13560-970, São Carlos, SP, Brazil}
\cortext[cor1]{Corresponding author}
\begin{abstract}
Weak measurements have an increasing number of applications in contemporary
quantum mechanics. They were originally described as a weak interaction
that slightly entangled the translational degrees of freedom of a
particle to its spin, yielding surprising results after post-selection.
That description often ignores the kinetic energy of the particle
and its movement in three dimensions. Here, we include these elements
and re-obtain the weak values within the context of collision theory
by two different approaches, and prove that the results are compatible
with each other and with the results from the traditional approach.
To provide a more complete description of the Stern-Gerlach apparatus, we use weak
vectors, a generalization of the weak values.
\end{abstract}

\end{frontmatter}{}

\section{Introduction}

New insight in science sometimes is gained when two different fields
of knowledge are recognized for the similarities they bear. Weak measurements,
first proposed in the 1980s by Aharonov et al. \cite{Aharonov1988},
and the much older formalism of collision theory are two such theories
containing striking parallels. In this article, we seek to survey
these similarities.

The concept of weak measurement, despite being quite more recent,
draws from sources as old as quantum mechanics itself, such as von
Neumann's formalism for the measurement. In his pioneering textbook,
von Neumann \cite{vN} dealt with the measurement process in quantum
mechanics as a unitary evolution governed by an interaction Hamiltonian
$H_{\mathrm{int}}$ between the main system and the measurement apparatus.
After this process, during which the two systems end up correlated,
occurss the reduction of the total state. At this point the terminology
presents a series of variations: according to \cite{ZurekRMP}, the
first step, when the unitary evolution occurs, is called pre-measurement,
while the collapse of the wave function is called measurement; Peres
\cite{Peres}, on the other hand, calls this global process an intervention,
divided in an interaction portion called measurement, and the collapse
which constitutes the output. This work by von Neumann was employed
as a starting point by many approaches to quantum mechanics, from
Everett's relative state formulation \cite{Everett,Wheeler,Dewitt}
to the decoherence theory \cite{JoosBook,SchlosshauerRMP,SchlosshauerBook,ZurekPT,ZurekPTL}
and pointer states \cite{Zeh1,Zeh2,ZurekPointer,BrasilEJP}, up to
analyses of the influence of the intensity of the interaction on the
probabilities associated with each result \cite{BrasilPRA,BrasilFP,BrasilEPJP}
and, finally, to weak values and measurements \cite{Aharonov1988,Ah1,Ah3,Duck}.

The concept of weak value emerged from the analysis made by Aharonov
et al. \cite{Ah4} of an ensemble both pre- and post-selected, in
an attempt to construct a new time-symmetric quantum theory. Even
though the concept of post-selection can initially sound uncanny \cite{Leggett,PeresPRL,Ah5},
we must highlight that we are not dealing with an isolated system,
but actually with two interacting Hilbert spaces, only one of which
suffers post-selection \textSFx{} a clear exposition of this is given
by Duck et al. \cite{Duck}. Between the pre- and the post-selection,
the intensity of the interaction $H_{\mathrm{int}}$ is made weak
enough so the higher-order terms in the power expansion of the time
evolution operator can be discarded, causing some ``surprising quantum
effects'' \cite{Ah1}. Namely, the weak value can surpass the limits
defined by the eigenvalue spectrum of the observable \cite{Aharonov1988,Duck}
and, as the weak values result from a weak interaction, the state
of the measured system remains practically unchanged \textSFx{} that
is, even though we are talking about ``measurement,'' there is no
collapse of the wave function.

In short, determining the weak value takes three measurements: (1)
the preparation of the initial state \textemdash{} or pre-selection
\textemdash{} which can be performed using a regular measurement;
(2) the weak measurement proper (a weak interaction that barely perturbs
the state of the system); and (3) the regular projective measurement,
performed in the post-selection stage. A collision experiment \cite{Weinberg,GellMann,Merzbacher}
is very similar to a weak measurement: there is an initial preparation
of the state that will be collided, the parts of the system that will
collide approach and interact during a finite time, and again depart,
until the point when, asymptotically, cease to interact, and, finally,
are detected by an apparatus that performs the final measurement,
or post-selection. The differential cross section, found by means
of a collisional experiment, is given in terms of the scattering matrix
\cite{GellMann,Moller,Wheeler2}.

A relationship between the scattering matrix and the weak values has
been established in \cite{Solli}. However, the authors treated the
scattering matrix as a response function of the system, exemplifying
their set-up by the construction of an effective scattering matrix
applied to an optical experiment using bi-refringent crystals.

In this work, we propose to establish a more specific parallel with
the formal collision theory, and apply it, aiming to create the possibility
of actual experimental set-ups with realistic collision, as in the
case of ultracold atomic collisions \cite{Williams}. Other controversial
questions about the weak value \cite{Hu,Vaidman,Kastner,Svensson,Ah6}
will not be dealt with here and will not be necessary for what we
propose to prove here. As recent reviews on the subject suggest \cite{Ah6,Ah7,Dressel},
the concept of weak value has become less abstract, and now can be
built from classical statistics \cite{Ferrie}. Worth of mention are
the proposal to use weak measurements to protect the state with finite-time
measurements \cite{BrasilEPJP}, and important experimental implementations,
for example, in optical interferometry to analyse the classical two-slit
experiment \cite{Danan}, and in condensed matter \cite{Acosta,Blok,Ibarcq}.

In Secs. 2 and 3, we will revise the concepts of weak of measurements
and collision theory, respectively, which we will apply concomitantly
in Sec. 4. Conclusions are presented in Sec. 5.

\section{Review of Weak Values}

Following the original article about weak measurements by Aharonov
et al. \cite{Aharonov1988,Ah3} and the alternative description by
I. M. Duck et al. \cite{Duck}, in this section we will consider an
interaction that entangles two quantum subsystems, one in a continuous
Hilbert space and the other in a discrete Hilbert space. We will treat
these two subsystems as a position (represented by the state vector
$\left|\Phi\right\rangle $) and a spin (represented by$\left|\chi\right\rangle $)
of a particle, just like in the Stern-Gerlach experiment. Initially,
we will assume they are not entangled to each other:

\begin{equation}
\left|\Psi_{\mathrm{i}}\right\rangle =\left|\Phi_{\mathrm{i}}\right\rangle \otimes\left|\chi_{\mathrm{i}}\right\rangle ,\label{PsiInit}
\end{equation}
where the subscript $\mathrm{i}$ indicates that this is the initial
state. The atom will be considered to have $1/2$ spin, so that any
observable of $\left|\chi\right\rangle $ will be proportional to
a linear combination of the Pauli matrices, $\sigma_{\mathbf{\hat{m}}}\equiv\boldsymbol{\sigma}\cdot\mathbf{\hat{m}}$,
where $\hat{\mathbf{m}}$ is a real unit vector and $\boldsymbol{\sigma}\equiv\sigma_{x}\mathbf{\hat{x}}+\sigma_{y}\mathbf{\hat{y}}+\sigma_{z}\mathbf{\hat{z}}$.

Throughout this work, we will represent the observables by capital
letters to differentiate them from their eigenvalues, which will be
written in lower case. Hence, we will represent the position operator
by $\mathbf{R}$ and its associated momentum observable by $\mathbf{P}$
. Their corresponding eigenstates are $\left|\mathbf{r}\right\rangle $
and $\left|\mathbf{p}\right\rangle $ with eigenvalues $\mathbf{r}$
and $\mathbf{p}$, respectively. The circumflex will be reserved to
unit vectors that are not operators, also written in lower case, as
in $\mathbf{\hat{x}}$.

Using this notation, the initial state of the translational degree
of freedom can be written as

\begin{equation}
\left|\Phi_{\mathrm{i}}\right\rangle =\int_{V_{r}^{\infty}}\mathrm{d}^{3}r\;\phi_{\mathrm{i}}\left(\mathbf{r}\right)\left|\mathbf{r}\right\rangle =\int_{V_{p}^{\infty}}\mathrm{d}^{3}p\;\varphi_{\mathrm{i}}\left(\mathbf{p}\right)\left|\mathbf{p}\right\rangle ,\label{PhiInit}
\end{equation}
where $V_{r}^{\infty}$ and $V_{p}^{\infty}$ indicate that the integrals
are performed in the entire three-dimensional spaces of positions
and momenta, respectively. The eigenstates of $\sigma_{\mathbf{\hat{m}}}$
are represented by the orthonormal set $\left\{ \left|0\right\rangle _{\mathbf{\hat{m}}},\left|1\right\rangle _{\mathbf{\hat{m}}}\right\} $,
so that:

\[
\sigma_{\mathbf{\hat{m}}}\left|s\right\rangle _{\mathbf{\hat{m}}}=\left(-1\right)^{s}\left|s\right\rangle _{\mathbf{\hat{m}}}.
\]
As we are assuming an initial state where spin and translation coordinate
are uncorrelated, the initial state of the former can be expanded
in terms of the eigenbasis of $\sigma_{\mathbf{\hat{m}}}$ by using
complex coefficients $\chi_{\mathrm{i}}^{\left(s\right)}$:

\begin{equation}
\left|\chi_{\mathrm{i}}\right\rangle =\sum_{s=0}^{1}\chi_{\mathrm{i}}^{\left(s\right)}\left|s\right\rangle _{\mathbf{\hat{m}}},\quad\sum_{s=0}^{1}\left|\chi_{\mathrm{i}}^{\left(s\right)}\right|^{2}=1.\label{CHIinit}
\end{equation}

The interaction Hamiltonian between the spin and the translational
degree of freedom that we will employ here is analogous to the one
proposed by von Neumann \cite{vN} for the interaction between a pointer
and some system it measures, before the collapse:

\begin{equation}
H_{\mathrm{int}}\left(t\right)=-\hbar\eta g\left(t\right)\mathbf{R}\cdot\overleftrightarrow{\mathrm{H}}\cdot\boldsymbol{\sigma}.\label{Hint}
\end{equation}
where $\eta$ is a real positive constant with dimension of inverse
position and time \textendash{} which we can consider small when we
want to study a weak measurement \textendash{} and $g\left(t\right)$
is a positive dimensionless function with compact support. This function
determines the duration of the interaction and we will assume that
its support falls entirely within the period of time elapsed between
pre- and post-selection. The tensor $\overleftrightarrow{\mathrm{H}}$
links the position and spin operators, and is often chosen as $\overleftrightarrow{\mathrm{H}}=\mathbf{\hat{z}}\mathbf{\hat{z}}$
in standard descriptions of the von Neumann measurement. This choice
makes it look like this tensor is irrelevant, but, as we will see
in Sec. 4, this common choice is not physically justified, so it will
be expedient to introduce it from the start.

The final state at $t=\tau_{\mathrm{f}}$, starting from an initial
state $\left|\Psi_{\mathrm{i}}\right\rangle $ at $t=-\tau_{\mathrm{i}}$,
will be the result of applying to Eq. (\ref{PsiInit}) the time evolution
operator derived from the interaction Hamiltonian from Eq. (\ref{Hint}):

\begin{equation}
\left|\Psi\left(\tau_{\mathrm{f}}\right)\right\rangle =\exp\left\{ -\frac{i}{\hbar}\int_{-\tau_{\mathrm{i}}}^{\tau_{\mathrm{f}}}\mathrm{d}t\:H_{\mathrm{int}}\left(t\right)\right\} \left|\Psi_{\mathrm{i}}\right\rangle =\exp\left\{ i\eta T\:\mathbf{R}\cdot\overleftrightarrow{\mathrm{H}}\cdot\boldsymbol{\sigma}\right\} \left|\Psi_{\mathrm{i}}\right\rangle .\label{PsiF}
\end{equation}
The exponential with the integral in Eq. (\ref{PsiF}) can be written
because, as seen from Eq. (\ref{Hint}), the interaction Hamiltonian
always commutes with itself at any instant of time ($\left[H_{\mathrm{int}}\left(t\right),H_{\mathrm{int}}\left(t^{\prime}\right)\right]=0$
for any $t$, $t^{\prime}$). Here, we are calling the normalization
of the function $g\left(t\right)$ as $T\equiv\int_{-\infty}^{\infty}\mathrm{d}t\:g\left(t\right)$,
which is equivalent to the time duration of the interaction if $g\left(t\right)$
is unit in the interval of its support.

Suppose we choose $\overleftrightarrow{\mathrm{H}}=\mathbf{\hat{z}}\mathbf{\hat{m}}$
and the initial state of the spin is $\left|s\right\rangle _{\mathbf{\hat{m}}}$.
Then, there will be a shift of $\pm\hbar\eta T$ in the wave function
$\varphi_{\mathrm{i}}\left(\mathbf{p}\right)$ from Eq. (\ref{PhiInit}),
because the exponential of a derivative acts as a Taylor series:

\begin{equation}
e^{i\eta Ts\mathbf{R}\cdot\mathbf{\hat{z}}}\varphi_{\mathrm{i}}\left(\mathbf{p}\right)\left|\mathbf{p}\right\rangle =\sum_{n=0}^{\infty}\frac{1}{n!}\left[\left(p_{z}-\hbar\eta Ts\right)-p_{z}\right]^{n}\frac{\partial^{n}\varphi_{\mathrm{i}}}{\partial p_{z}^{n}}\left|\mathbf{p}\right\rangle =\varphi_{\mathrm{i}}\left(\mathbf{p}-\hbar\eta Ts\mathbf{\hat{z}}\right)\left|\mathbf{p}\right\rangle .\label{ExpTaylor}
\end{equation}
In Eq. (\ref{ExpTaylor}) above, we used the fact that the position
operator projected in the basis of momenta acts as $\mathbf{R}\tilde{\phi}_{\mathrm{i}}\left(\mathbf{p}\right)\left|\mathbf{p}\right\rangle =i\hbar\nabla_{p}\varphi_{\mathrm{i}}\left(\mathbf{p}\right)\left|\mathbf{p}\right\rangle $.

When applied to any given initial state from Eqs. (\ref{PhiInit})
and (\ref{CHIinit}), the time-evolution operator from Eq. (\ref{PsiF})
becomes:

\begin{equation}
\left|\Psi\left(\tau_{\mathrm{f}}\right)\right\rangle =\sum_{s=0}^{1}\chi_{\mathrm{i}}^{\left(s\right)}\int_{V_{p}^{\infty}}\mathrm{d}^{3}p\:\varphi_{\mathrm{i}}\left(\mathbf{p}-\hbar\eta T\left(-1\right)^{s}\mathbf{\hat{z}}\right)\left|\mathbf{p}\right\rangle \otimes\left|s\right\rangle _{\mathbf{\hat{m}}}.\label{Von Neumann measurement}
\end{equation}
Whether the wave function is shifted to the left or to the right depends
on the value of $s$. If the shift is greater than the variance of
the distribution, measuring the momentum gives a good estimate of
the value of $s$. This is illustrated in Fig. 1.

\begin{figure}
\includegraphics[width=0.49\textwidth]{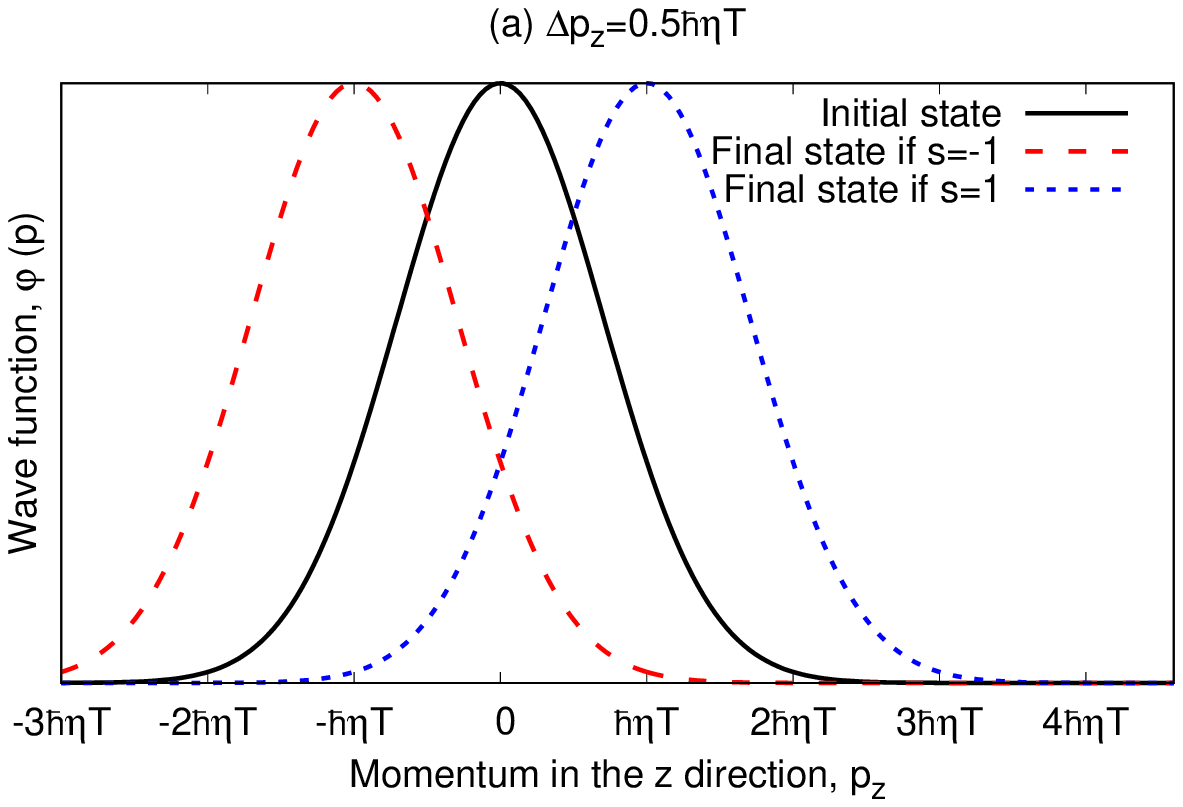}\includegraphics[width=0.49\textwidth]{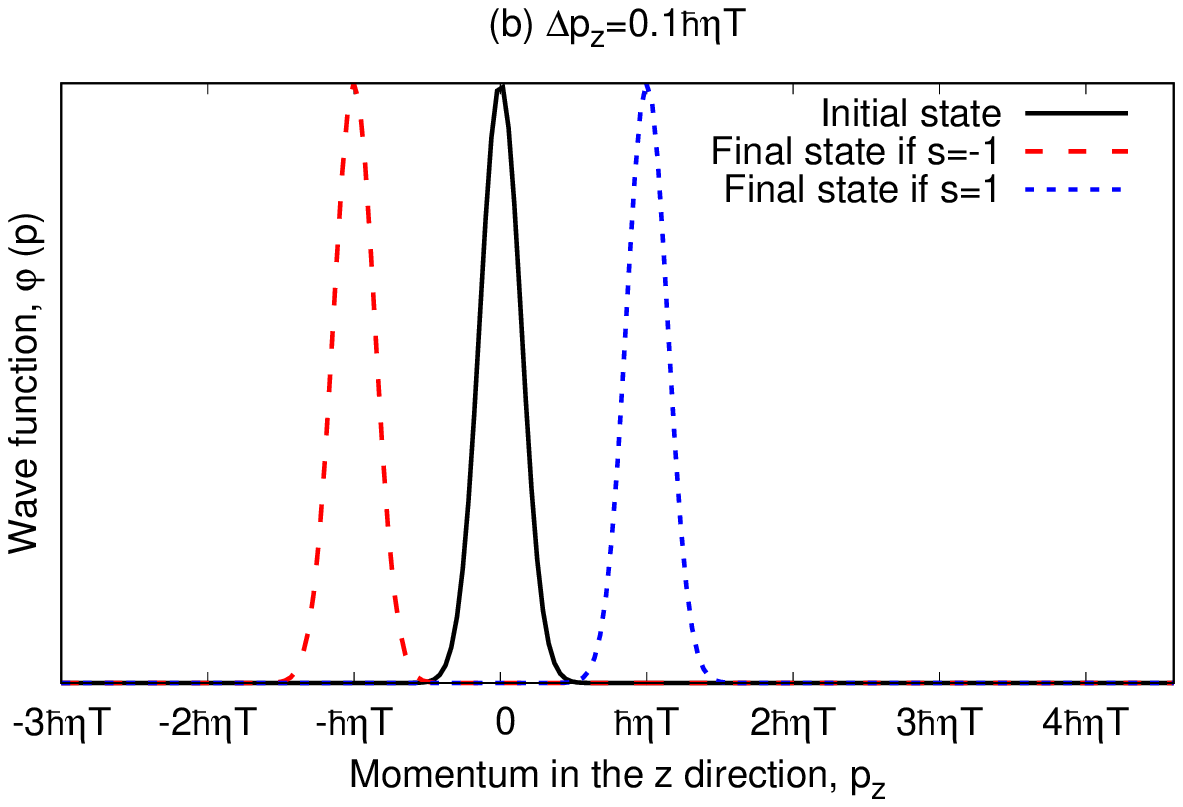}

\caption{Displacement of a Gaussian wave function in the space of momenta after
a von Neumann measurement. The direction of the displacement indicates
the result of a measurement of $\sigma_{\mathbf{\hat{m}}}$. If the
dispersion is small compared to the displacement of the wave function
\textemdash{} as in (b) \textemdash{} the result of the measurement
of the momentum $p_{z}$ will always be close the value of the spin
multiplied by $\hbar\eta T$.}
\end{figure}

In the formalism presented by Aharonov et al. \cite{Ah1,Aharonov1988,Ah3},
besides making the measurement weak by taking a small value for $\eta$
{[}as in Fig. 1(a){]}, the system is also subject to a post-selection.
Just like pre-selection determines the initial state of the system
$\left|\Psi_{\mathrm{i}}\right\rangle $, post-selection defines part
of the final state $\left|\Psi_{\mathrm{f}}\right\rangle $, by discarding
the result unless if it gives a different value. Discarding the entire
system unless it is exactly what we want can hardly be useful in experiments,
so we discard it unless the final state of only the discrete part,
the spin, is measured in a certain state $\left|\chi_{\mathrm{f}}\right\rangle $.

The final state of the translational degree of freedom after the spin
has been determined to be at $\left|\chi_{\mathrm{f}}\right\rangle $
can be found by projecting the final joint state given in Eq. (\ref{PsiF})
into it:

\begin{equation}
\mathrm{rel}_{\left|\chi_{\mathrm{f}}\right\rangle }\left|\Psi\left(\tau_{\mathrm{f}}\right)\right\rangle \equiv C\left\langle \chi_{\mathrm{f}}\right|\left.\Psi\left(\tau_{\mathrm{f}}\right)\right\rangle ,\label{postselection}
\end{equation}
where we are using a notation that identifies the post-selected state
with a relative state \cite{BrasilEJP}, and where the constant $C\equiv\left|\left\langle \chi_{\mathrm{f}}\right|\left.\Psi\left(\tau_{\mathrm{f}}\right)\right\rangle \right|^{-1}$
is required to normalize the state vector.

Using the power expansion from Eq. (\ref{ExpTaylor}), the post-selected
state from Eq. (\ref{postselection}) becomes:

\begin{eqnarray}
\mathrm{rel}_{\left|\chi_{\mathrm{f}}\right\rangle }\left|\Psi\left(\tau_{\mathrm{f}}\right)\right\rangle  & = & C\left\langle \chi_{\mathrm{f}}\right|\left.\chi_{\mathrm{i}}\right\rangle \left[\sum_{n=0}^{\infty}\frac{1}{n!}\left(i\eta T\right)^{n}\left(\mathbf{R}\cdot\overleftrightarrow{\mathrm{H}}\right)^{\otimes n}\cdot\boldsymbol{\sigma}_{\mathrm{w}}^{\left(n\right)}\right]\left|\Phi_{\mathrm{i}}\right\rangle ,\label{postSelectedExpansion}
\end{eqnarray}
where the tensor power $A^{\otimes n}$ represents $n$ tensor products
of the same operator $A$. Generalizing Aharonov et al. \cite{Ah1,Ah3,Duck},
we defined the $n$th weak tensor of the vector operator $\boldsymbol{\sigma}$
as:

\begin{eqnarray}
\boldsymbol{\sigma}_{\mathrm{w}}^{\left(n\right)} & \equiv & \frac{\left\langle \chi_{\mathrm{f}}\right|\boldsymbol{\sigma}^{\otimes n}\left|\chi_{\mathrm{i}}\right\rangle }{\left\langle \chi_{\mathrm{f}}\right|\left.\chi_{\mathrm{i}}\right\rangle }.\label{weak tensor}
\end{eqnarray}
If $\eta$ is small enough, the series in Eq. (\ref{postSelectedExpansion})
can be simplified by dropping higher-order terms. The only weak tensor
to be kept then is $\boldsymbol{\sigma}_{\mathrm{w}}\equiv\boldsymbol{\sigma}_{\mathrm{w}}^{\left(1\right)}$
which is known simply as weak vector \cite{Ah3}.
It is important to emphasize that the weak vector $\boldsymbol{\sigma}_{\mathrm{w}}$
(unlike $\boldsymbol{\sigma}$) is a vector in the three-dimensional
Cartesian space, not an operator in Hilbert space. Hence:

\begin{equation}
\mathrm{rel}_{\left|\chi_{\mathrm{f}}\right\rangle }\left|\Psi\left(\tau_{\mathrm{f}}\right)\right\rangle =C\left\langle \chi_{\mathrm{f}}\right|\left.\chi_{\mathrm{i}}\right\rangle e^{i\eta T\mathbf{R}\cdot\overleftrightarrow{\mathrm{H}}\cdot\boldsymbol{\sigma}_{\mathrm{w}}}\left|\Phi_{\mathrm{i}}\right\rangle +O\left(\eta^{2}\right).\label{exponential weak vector}
\end{equation}

If we write the initial state $\left|\Phi_{\mathrm{i}}\right\rangle $
explicitly as in Eq. (\ref{PhiInit}), we can express the post-selected
state as:

\[
\mathrm{rel}_{\left|\chi_{\mathrm{f}}\right\rangle }\left|\Psi\left(\tau_{\mathrm{f}}\right)\right\rangle =C\left\langle \chi_{\mathrm{f}}\right|\left.\chi_{\mathrm{i}}\right\rangle \int_{V_{p}^{\infty}}\mathrm{d}^{3}p\:\exp\left\{ -\hbar\eta T\boldsymbol{\nabla}_{p}\cdot\overleftrightarrow{\mathrm{H}}\cdot\boldsymbol{\sigma}_{\mathrm{w}}\right\} \varphi_{\mathrm{i}}\left(\mathbf{p}\right)\left|\mathbf{p}\right\rangle +O\left(\eta^{2}\right).
\]
Using the Taylor series to once again simplify the exponential of
the derivative, as we did in Eq. (\ref{ExpTaylor}), we can write
that the weak measurement displaced the original wave function by
an amount proportional to the weak vector $\boldsymbol{\sigma}_{\mathrm{w}}$:

\begin{equation}
\mathrm{rel}_{\left|\chi_{\mathrm{f}}\right\rangle }\left|\Psi\left(\tau_{\mathrm{f}}\right)\right\rangle =C\left\langle \chi_{\mathrm{f}}\right|\left.\chi_{\mathrm{i}}\right\rangle \int_{V_{p}^{\infty}}\mathrm{d}^{3}p\:\varphi_{\mathrm{i}}\left(\mathbf{p}-\hbar\eta T\overleftrightarrow{\mathrm{H}}\cdot\boldsymbol{\sigma}_{\mathrm{w}}\right)\left|\mathbf{p}\right\rangle +O\left(\eta^{2}\right).\label{weak measurement}
\end{equation}

If, as in Eq. (\ref{Von Neumann measurement}), we identify the displacement
of the wave function with the value of a measurement of the spin,
then the weak vectors will contain the results of such measurements.
For example, a shift of $\hbar\eta T\mathbf{\hat{m}}\cdot\overleftrightarrow{\mathrm{H}}\cdot\boldsymbol{\sigma}_{\mathrm{w}}$
in the $\mathbf{\hat{m}}$ direction represents a measurement of $\mathbf{\hat{m}}\cdot\overleftrightarrow{\mathrm{H}}\cdot\boldsymbol{\sigma}_{\mathrm{w}}$
for the observable $\mathbf{\hat{m}}\cdot\overleftrightarrow{\mathrm{H}}\cdot\boldsymbol{\sigma}$.
This can lead to results not possible in ordinary measurements.

\begin{figure}
\includegraphics[width=0.49\textwidth]{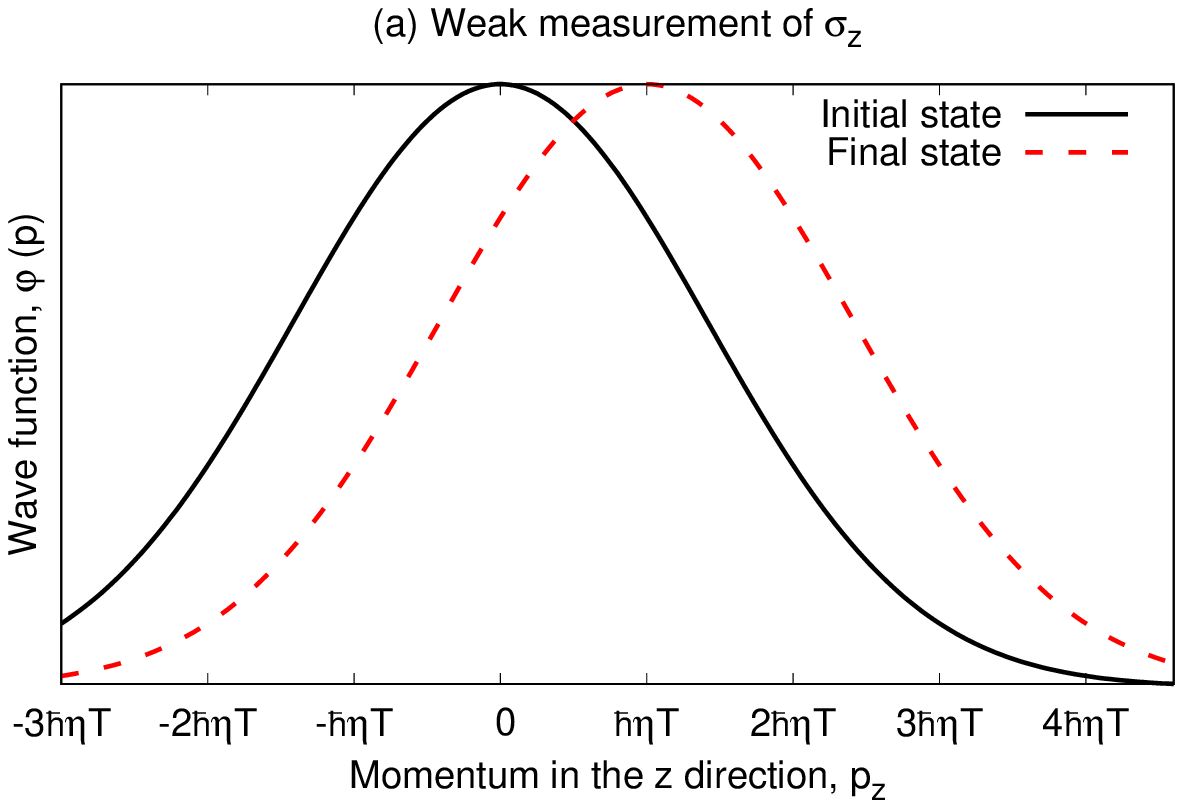}\includegraphics[width=0.49\textwidth]{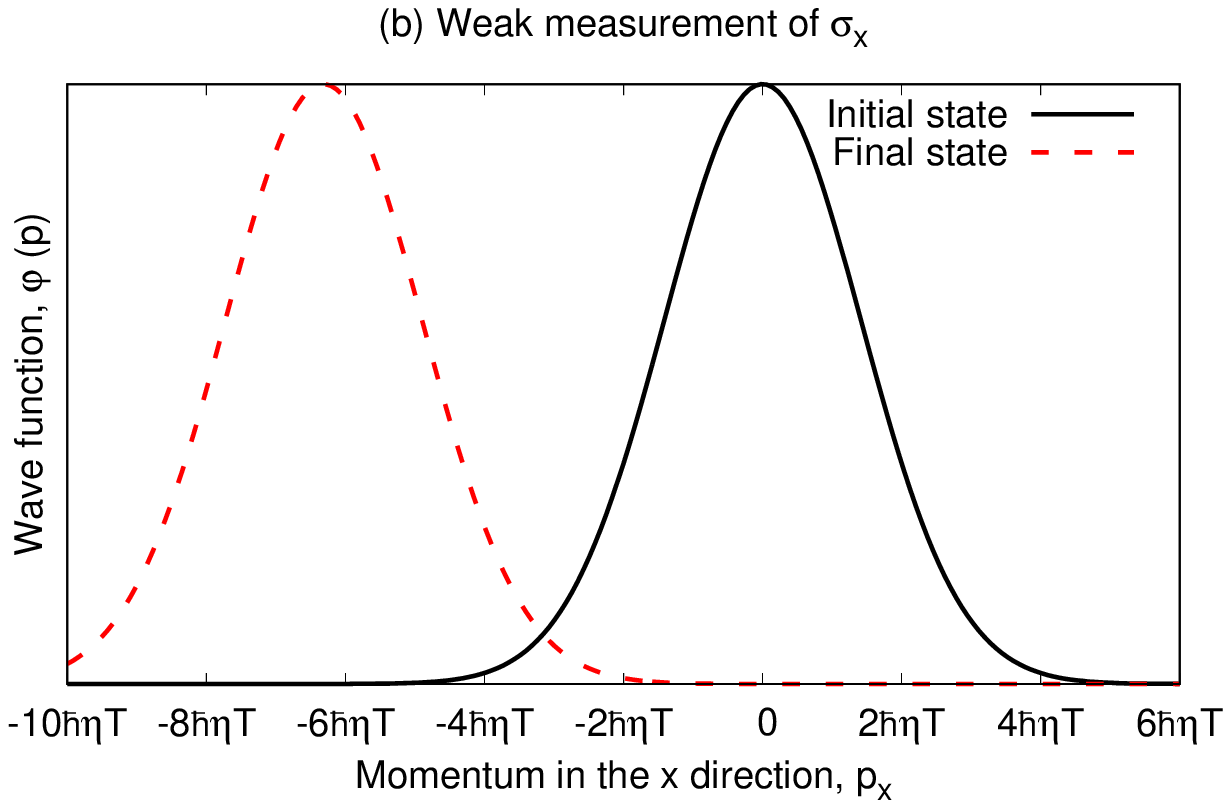}

\includegraphics[width=0.5\textwidth]{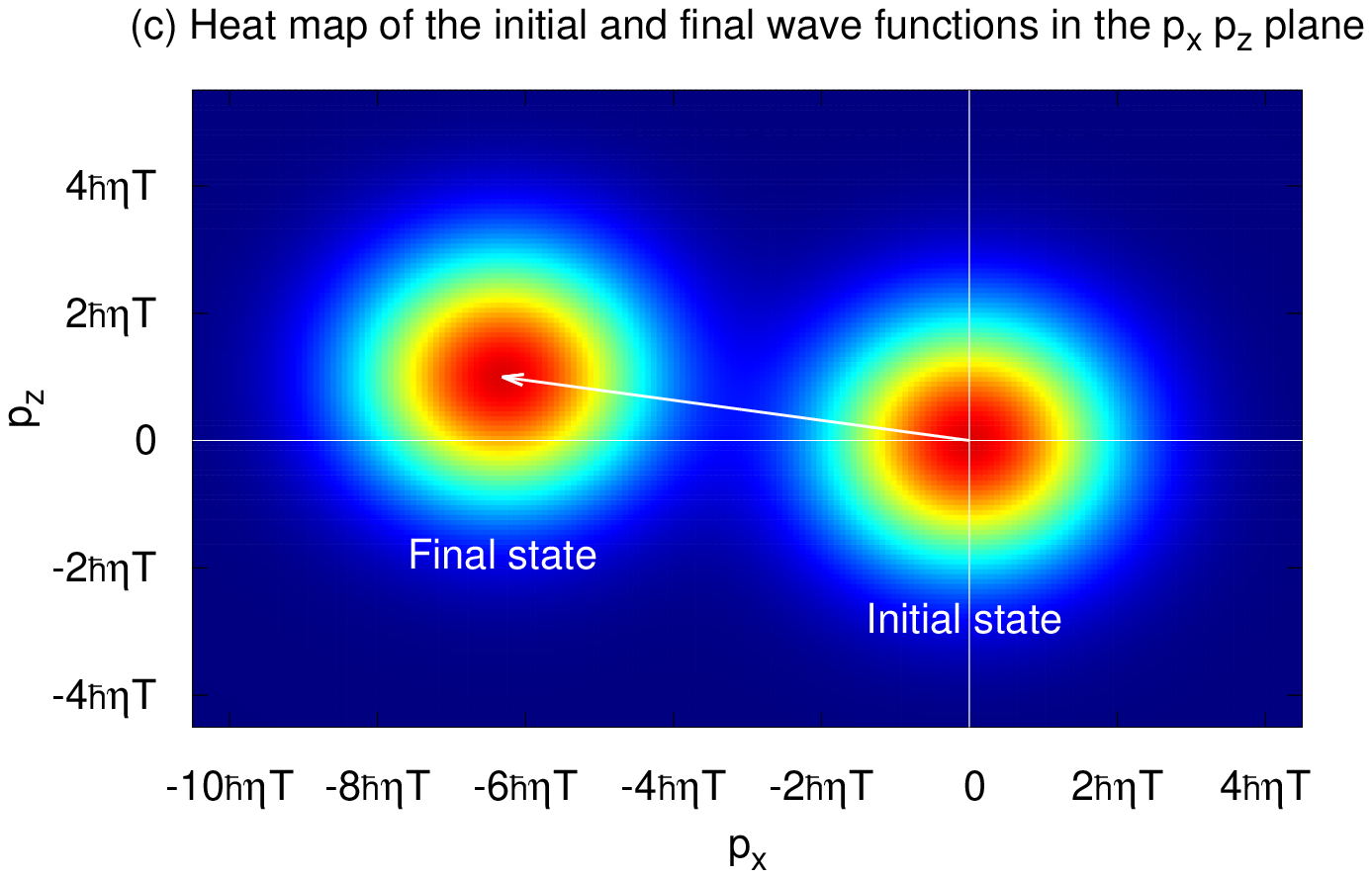}

\caption{Displacements of the wave function in the momentum space for a (a)
a weak measurement of $\sigma_{z}$ ; and (b) a weak measurement of
$\sigma_{x}$, for $\theta=9\pi/10$. The first one lies within bounds
of a normal measurement, while the second one causes a displacement
that goes far beyond it. The information of these two graphs is contained
in the bidimensional heatmap shown in (c), where the displacement
$\mathbf{\hat{z}}-\tan\left(9\pi/20\right)\mathbf{\hat{x}}$ of the
wave function due to the weak vector is shown explicitly in an arrow.}
\end{figure}

This is illustrated in  2, where we consider the more realistic case
(as will be explained in Sec. 4) where $\overleftrightarrow{\mathrm{H}}=\mathbf{\hat{z}}\mathbf{\hat{z}}-\mathbf{\hat{x}}\mathbf{\hat{x}}$.
We choose an initially Gaussian wave function centered at the origin
of the momentum space, with a post-selected state $\left|\chi_{\mathrm{f}}\right\rangle =\left|0\right\rangle _{\mathbf{\hat{z}}}$
and a pre-selected state $\left|\chi_{\mathrm{i}}\right\rangle =\left|0\right\rangle _{\mathbf{\hat{m}}}$,
with $\mathbf{\hat{m}}=\cos\theta\mathbf{\hat{z}}+\sin\theta\mathbf{\hat{x}}$
so that $\left|\chi_{\mathrm{i}}\right\rangle =\cos\left(\theta/2\right)\left|0\right\rangle _{\mathbf{\hat{z}}}+\sin\left(\theta/2\right)\left|1\right\rangle _{\mathbf{\hat{z}}}$.
Then, the weak vector becomes $\boldsymbol{\sigma}_{\mathrm{w}}=\mathbf{\hat{z}}+\tan\left(\theta/2\right)\mathbf{\hat{x}}-i\tan\left(\theta/2\right)\mathbf{\hat{y}}$,
so that the initial wave function is displaced by $\hbar\eta T\mathbf{\hat{z}}-\hbar\eta T\tan\left(\theta/2\right)\mathbf{\hat{x}}$.
This means that a measurement of $\sigma_{z}$ yields a result $s=1$,
which is the expected for a post-selection of the eigenstate $\left|0\right\rangle _{\mathbf{\hat{z}}}$.
However, the measurement of $\sigma_{x}$ can yield an unphysical
value of $\left|s\right|>1$, displaying an example of the \textquotedblleft surprising
effects\textquotedblright{} \cite{Ah1} of the weak measurement.

As we have seen in this section, the concept of weak vector, which
can be ignored in simpler description of the weak measurement, encapsulates
all the information necessary to describe different weak measurements
and will prove itself useful in describing more realistic experiments
in Sec. 4. However, we shall first review collision theory, which
will later be used to re-derive the results of weak measurements.

\section{Review of Collision Theory}

In the previous section, we considered only a Hamiltonian that coupled
translational and spin degrees of freedom, but entirely neglected
the kinetic energy of the atomic beam. In this section, we will review
some concepts of atomic collision theory and, as such, consider explicitly
the kinetic energy contribution to the Hamiltonian, $H_{0}$. We will
also consider that the atomic beam is described by a wave packet that
approaches a target in whose neighborhood some generic interaction
Hamiltonian $H_{\mathrm{int}}$ acts on the system.

In the beginning and in the end, the beam is too distant to interact
with the target, and the state will evolve according to an unperturbed
Hamiltonian $H_{0}$. Supposing the particles of the beam do not interact
with each other, for each one of them,

\begin{equation}
H_{0}=\frac{P^{2}}{2m},\label{H0}
\end{equation}
where $P$ is the observable corresponding to the absolute value of
the linear momentum ($P^{2}=P_{x}^{2}+P_{y}^{2}+P_{z}^{2}$), and
$m$ is the particle's mass. Clearly, $H_{0}$ does not entangle the
spin to the translational degrees of freedom and so $\left|\chi\right\rangle $
can be ignored while $H_{\mathrm{int}}$ is not acting. It may be
convenient to write the initial state from Eq. (\ref{PhiInit}) in
terms of the eigenstates of $H_{0}$. These states are proportional
to the eigenstates of the momentum, $\left|\mathbf{p}\right\rangle $,
the difference being their normalization.

The eigenstates with the same absolute value $p$ of the momentum
\textemdash{} and hence energy eigenvalue $E=p^{2}/2m$ \textemdash{}
will be degenerate, only differing in the direction $\mathbf{\hat{p}}$
of the momentum, and so will be represented by $\left|\Phi_{\mathbf{\hat{p}}}\left(E\right)\right\rangle $.
While the normalization of the momentum is $\left\langle \mathbf{p^{\prime}}\right.\left|\mathbf{p}\right\rangle =\delta^{\left(3\right)}\left(\mathbf{p}^{\prime}-\mathbf{p}\right)$,
the normalization of the eigenstates of the kinetic energy is $\left\langle \Phi_{\mathbf{\hat{p}^{\prime}}}\left(E^{\prime}\right)\right.\left|\Phi_{\mathbf{\hat{p}}}\left(E\right)\right\rangle =\delta\left(E^{\prime}-E\right)\delta^{\left(2\right)}\left(\hat{\mathbf{p}}^{\prime}-\mathbf{\hat{p}}\right)$.
Therefore, the difference in normalization can be found from Dirac
delta's identities:

\begin{equation}
\left|\Phi_{\mathbf{\hat{p}}}\left(E=\frac{p^{2}}{2m}\right)\right\rangle =\sqrt{pm}\left|\mathbf{p}\right\rangle .\label{eigenstatesH0}
\end{equation}

We will choose the origin of our coordinate system at the center of
the scattering target. Therefore, close to $\mathbf{r}=\mathbf{0}$,
the total Hamiltonian is $H=H_{0}+H_{\mathrm{int}}$, whose eigenstates
will be represented by $\left|\Psi\left(E\right)\right\rangle $:

\begin{eqnarray}
H\left|\Psi\left(E\right)\right\rangle  & = & E\left|\Psi\left(E\right)\right\rangle .\label{eigenstatesH}
\end{eqnarray}
As the $H_{\mathrm{int}}$ may be capable of entangling the spin and
the translational degree of freedom, $\left|\Psi\left(E\right)\right\rangle $
may no longer be described solely in terms of the translational coordinates.
Considering the interaction Hamiltonian as a perturbation, we may
explicitly write the small constant $\eta$ in front of it in $H=H_{0}+\eta H_{\mathrm{int}}$.
When $\eta\to0$, we expect the eigenvectors of $H$ to be reduced
to the eigenvectors of $H_{0}$. Defining $\left|\Phi_{\mathbf{\hat{p}},s}\left(E\right)\right\rangle \equiv\left|\Phi_{\mathbf{\hat{p}}}\left(E\right)\right\rangle \otimes\left|s\right\rangle _{\mathbf{\hat{m}}}$
(where the direction $\mathbf{\hat{m}}$ of the spin can be chosen
according to convenience) we see that we can employ the same subscripts
used to identify the eigenstates of $H_{0}$ to identify the eigenstates
of $H$, because the two sets are related by:

\[
\lim_{\eta\to0}\left|\Psi_{\mathbf{\hat{p}},s}\left(E\right)\right\rangle =\left|\Phi_{\mathbf{\hat{p}},s}\left(E\right)\right\rangle .
\]

Our main objective in a scattering problem will be to find the probability
that a system initially prepared in the state $\left|\Psi_{\mathrm{i}}\right\rangle $
at $t=-\tau$ will be detected at the state $\left|\Psi_{\mathrm{f}}\right\rangle $
at $t=\tau$ (where $\tau>0$). The probability amplitude $P_{\mathrm{if}}$
of this transition for a time-independent Hamiltonian $H$ is $\left\langle \Psi_{\mathrm{f}}\right|e^{-2iH\tau/\hbar}\left|\Psi_{\mathrm{i}}\right\rangle $.
Suppose we write the vector states $\left|\Psi_{\mathrm{i}}\right\rangle $
and $\left|\Psi_{\mathrm{f}}\right\rangle $ in an interaction picture,
incorporating the Hamiltonian term $H_{0}$ into them by writing $\left|\tilde{\Psi}_{\mathrm{i}}\right\rangle \equiv e^{-iH_{0}\tau/\hbar}\left|\Psi_{\mathrm{i}}\right\rangle $
and $\left|\tilde{\Psi}_{\mathrm{f}}\right\rangle \equiv e^{iH_{0}\tau/\hbar}\left|\Psi_{\mathrm{f}}\right\rangle $.
In this case, the probability amplitude $P_{\mathrm{if}}$ can be
written as:

\[
P_{\mathrm{if}}=\left\langle \tilde{\Psi}_{\mathrm{f}}\right|e^{iH_{0}\tau/\hbar}e^{-2iH\tau/\hbar}e^{iH_{0}\tau/\hbar}\left|\tilde{\Psi}_{\mathrm{i}}\right\rangle .
\]

In the limit when $\tau\to\infty$, we can define the Møller operators
$\Omega^{\pm}\equiv\lim_{\tau\to\mp\infty}e^{iH\tau/\hbar}e^{-iH_{0}\tau/\hbar}$
so we can write $P_{\mathrm{if}}$ as

\begin{equation}
P_{\mathrm{if}}=\left\langle \tilde{\Psi}_{\mathrm{f}}\right|\left[\Omega^{-}\right]^{\dagger}\Omega^{+}\left|\tilde{\Psi}_{\mathrm{i}}\right\rangle .\label{Probs}
\end{equation}
The operator $S\equiv\left[\Omega^{-}\right]^{\dagger}\Omega^{+}$
is often called S-matrix or scattering matrix.

We can write the initial and final states in Eq. (\ref{Probs}) as
a superposition of eigenstates of $H_{0}$:

\begin{align}
P_{\mathrm{if}} & =\sum_{s,s^{\prime}}\int_{0}^{\infty}\mathrm{d}E\int_{4\pi}\mathrm{d}\Omega\int_{0}^{\infty}\mathrm{d}E^{\prime}\int_{4\pi}\mathrm{d}\Omega^{\prime}\:\left\langle \tilde{\Psi}_{\mathrm{f}}\right|\left.\Phi_{\mathbf{\hat{p}}^{\prime},s^{\prime}}\left(E^{\prime}\right)\right\rangle \left\langle \Phi_{\mathbf{\hat{p}},s}\left(E\right)\right|\left.\tilde{\Psi}_{\mathrm{i}}\right\rangle \nonumber \\
 & \qquad\times\left\langle \Phi_{\mathbf{\hat{p}}^{\prime},s^{\prime}}\left(E^{\prime}\right)\right|S\left|\Phi_{\mathbf{\hat{p}},s}\left(E\right)\right\rangle .\label{Probs2}
\end{align}

The next step requires that we know how to calculate the S-matrix
element present in Eq. (\ref{Probs}). Here, it is convenient to use
the Lippmann-Schwinger equation \cite{Weinberg,LS1,LS2}:
\begin{eqnarray}
\left|\Psi_{\mathbf{\hat{p}},s}^{\pm}\left(E\right)\right\rangle  & = & \left|\Phi_{\mathbf{\hat{p}},s}\left(E\right)\right\rangle +\eta\lim_{\varepsilon\to0\pm}\frac{1}{E-H_{0}+i\varepsilon}H_{\mathrm{int}}\left|\Psi_{\mathbf{\hat{p}},s}^{\pm}\left(E\right)\right\rangle ,\label{LippmannSchwinger}
\end{eqnarray}
where the inverse operator can be understood as:

\[
\frac{1}{E-H_{0}+i\varepsilon}\equiv\sum_{s^{\prime}}\int_{0}^{\infty}\mathrm{d}E^{\prime}\;\frac{1}{E-E^{\prime}+i\varepsilon}\int_{4\pi}\mathrm{d}\Omega^{\prime}\left|\Phi_{\mathbf{\hat{p}}^{\prime},s^{\prime}}\left(E^{\prime}\right)\right\rangle \left\langle \Phi_{\mathbf{\hat{p}}^{\prime},s^{\prime}}\left(E^{\prime}\right)\right|.
\]

The eigenvectors $\left|\Psi_{\mathbf{\hat{p}},s}^{\pm}\left(E\right)\right\rangle $
of $H$ are simply the Møller operators $\Omega^{\pm}$ applied to
the eigenvectors $\left|\Phi_{\mathbf{\hat{p}},s}\left(E\right)\right\rangle $
of $H_{0}$, because these operators evolve the system under $H_{0}$
either to infinity or minus infinity time and then evolve them back
to $t=0$ under the new Hamiltonian $H$. A more formal proof of this
is given in Appendix A. Therefore, the elements of the S-matrix can
be written in terms of these states:

\begin{equation}
\left\langle \Phi_{\mathbf{\hat{p}}^{\prime},s^{\prime}}\left(E^{\prime}\right)\right|S\left|\Phi_{\mathbf{\hat{p}},s}\left(E\right)\right\rangle =\left\langle \Psi_{\mathbf{\hat{p}}^{\prime},s^{\prime}}^{-}\left(E^{\prime}\right)\right.\left|\Psi_{\mathbf{\hat{p}},s}^{+}\left(E\right)\right\rangle .\label{SSandwich}
\end{equation}

Up to the first order in $\eta$, the product from Eq. (\ref{SSandwich})
can be written using Eq. (\ref{LippmannSchwinger}):

\begin{align*}
\left\langle \Psi_{\mathbf{\hat{p}}^{\prime},s^{\prime}}^{-}\left(E^{\prime}\right)\right.\left|\Psi_{\mathbf{\hat{p}},s}^{+}\left(E\right)\right\rangle  & =\left\langle \Phi_{\mathbf{\hat{p}}^{\prime},s^{\prime}}\left(E^{\prime}\right)\right|\left\{ 1-\eta\lim_{\varepsilon\to0+}\frac{2i\varepsilon}{\left(E-E^{\prime}\right)^{2}+\varepsilon^{2}}H_{\mathrm{int}}\right\} \left|\Phi_{\mathbf{\hat{p}},s}\left(E\right)\right\rangle \\
 & \quad+O\left(\eta^{2}\right).
\end{align*}

Using the fact that the limit results in the Dirac delta, because,
for any function $f\left(x\right)$,

\[
\lim_{\varepsilon\to0+}\int_{-\infty}^{\infty}\mathrm{d}x\frac{2i\varepsilon}{x^{2}+\varepsilon^{2}}f\left(x\right)=2\pi i\lim_{\varepsilon\to0+}\frac{2i\varepsilon}{2\left(i\varepsilon\right)}f\left(i\varepsilon\right)=2\pi i\int_{-\infty}^{\infty}\mathrm{d}x\:\delta\left(x\right)f\left(x\right),
\]
we find:

\begin{align}
\left\langle \Psi_{\mathbf{\hat{p}}^{\prime},s^{\prime}}^{-}\left(E^{\prime}\right)\right.\left|\Psi_{\mathbf{\hat{p}},s}^{+}\left(E\right)\right\rangle  & =\delta\left(E-E^{\prime}\right)\left\{ \delta^{\left(2\right)}\left(\mathbf{\hat{p}}^{\prime}-\mathbf{\hat{p}}\right)\delta_{s,s^{\prime}}-2\pi i\left\langle \Phi_{\mathbf{\hat{p}}^{\prime},s^{\prime}}\left(E^{\prime}\right)\right|\eta H_{\mathrm{int}}\left|\Phi_{\mathbf{\hat{p}},s}\left(E\right)\right\rangle \right\} \nonumber \\
 & \qquad+O\left(\eta^{2}\right).\label{inner product}
\end{align}

The inner product 
\[
T_{\mathbf{\hat{p}}^{\prime},s^{\prime};\mathbf{\hat{p}},s}\left(E^{\prime};E\right)\equiv\left\langle \Phi_{\mathbf{\hat{p}}^{\prime},s^{\prime}}\left(E^{\prime}\right)\right|\eta H_{\mathrm{int}}\left|\Psi_{\mathbf{\hat{p}},s}\left(E\right)\right\rangle =\left\langle \Phi_{\mathbf{\hat{p}}^{\prime},s^{\prime}}\left(E^{\prime}\right)\right|\eta H_{\mathrm{int}}\left|\Phi_{\mathbf{\hat{p}},s}\left(E\right)\right\rangle +O\left(\eta^{2}\right)
\]
is often called transition matrix or T-matrix.

Replacing Eq. (\ref{inner product}) at the probability amplitude
$P_{\mathrm{if}}$ from Eq. (\ref{Probs2}):

\begin{equation}
P_{\mathrm{if}}=\left\langle \tilde{\Psi}_{\mathrm{f}}\right|\left.\tilde{\Psi}_{\mathrm{i}}\right\rangle -2\pi i\eta\left\langle \tilde{\Psi}_{\mathrm{f}}\right|H_{\mathrm{int}}\delta\left(E-E^{\prime}\right)\left|\tilde{\Psi}_{\mathrm{i}}\right\rangle +O\left(\eta^{2}\right),\label{PifFinal}
\end{equation}
where the Dirac delta is a compact notation for

\[
H_{\mathrm{int}}\delta\left(E-E^{\prime}\right)=\sum_{s,s^{\prime}}\int_{0}^{\infty}\mathrm{d}E\int_{4\pi}\mathrm{d}\Omega\int_{4\pi}\mathrm{d}\Omega^{\prime}\left|\Phi_{\mathbf{\hat{p}}^{\prime},s^{\prime}}\left(E\right)\right\rangle \left\langle \Phi_{\mathbf{\hat{p}}^{\prime},s^{\prime}}\left(E\right)\right|H_{\mathrm{int}}\left|\Phi_{\mathbf{\hat{p}},s}\left(E\right)\right\rangle \left\langle \Phi_{\mathbf{\hat{p}},s}\left(E\right)\right|,
\]
which, according to Eq. (\ref{eigenstatesH0}), can be written also
as:

\[
H_{\mathrm{int}}\delta\left(E-E^{\prime}\right)=\int_{V_{p}^{\infty}}\mathrm{d}^{3}p^{\prime}\int_{V_{p}^{\infty}}\mathrm{d}^{3}p\:\delta\left(\frac{p^{2}}{2m}-\frac{p^{\prime2}}{2m}\right)\left|\mathbf{p}^{\prime}\right\rangle \left\langle \mathbf{p}^{\prime}\right|H_{\mathrm{int}}\left|\mathbf{p}\right\rangle \left\langle \mathbf{p}\right|.
\]

All we need now to determine the probability amplitudes are the initial
and final states and the interaction Hamiltonian. We will use this
formalism applied to the Stern-Gerlach experiment and to weak measurements
in the next section.

\section{The Weak Measurement as a Collision}

In this section we will describe a weak measurement that has an interaction
Hamiltonian similar to that employed in Sec. 2 to which we will apply
the collisional formalism from Sec. 3. The interaction term of the
Hamiltosnian is the result of the Stern-Gerlach interaction between
the magnetic dipole moment of the spin-$1/2$, given by $g_{\mathrm{S}}\mu_{\mathrm{B}}\boldsymbol{\sigma}/2$
\textemdash{} where $g_{\mathrm{S}}\approx2.00231930419922$ and $\mu_{\mathrm{B}}$
is Bohr's magneton \textemdash{} and an external magnetic field $\mathbf{B}\left(\mathbf{r}\right)$
that varies with the position:

\begin{equation}
H_{\mathrm{SG}}=\frac{1}{2}g_{S}\mu_{\mathrm{B}}\boldsymbol{\sigma}\cdot\mathbf{B}\left(\mathbf{R}\right),\label{HSG}
\end{equation}
where $\mathbf{R}$ is the position operator.

As we are interested in coupling the spin with the translational degrees
of freedom of the beam, it is necessary that $\mathbf{B}$ be linearly
dependent on the position operators, like Eq. (\ref{Hint}). This
can be approximately obtained, for example, if the field is zero at
the origin and the beam does not go far from it, so that the magnetic
field can be expanded as a power series such as $\mathbf{B}\left(x,z\right)\approx\mathbf{\hat{x}}x\partial_{x}B\left(x=0,z=0\right)+\mathbf{\hat{z}}z\partial_{x}B\left(x=0,z=0\right)$.
Also, according to the symmetry of the apparatus, we can consider
that the magnetic field is confined to the plane perpendicular to
the direction of the beam. Considering it to be the $y$ axis, we
restrict the field $\mathbf{B}$ to the $xz$ plane and, likewise,
can consider that it depends only on the $x$ and $z$ coordinates:

\[
\mathbf{B}\left(\mathbf{r}\right)=\mathbf{\hat{x}}\left(B_{xx}x+B_{xz}z\right)+\mathbf{\hat{z}}\left(B_{zx}x+B_{zz}z\right),
\]
where $B_{xx}$, $B_{xz}$, $B_{zx}$, $B_{zz}$ are constant scalars.

As the part of the Stern-Gerlach apparatus where the beam propagates
has no currents, charges nor time-varying fields, $\mathbf{B}\left(\mathbf{r}\right)$
must obey the Maxwell equations $\boldsymbol{\nabla}\cdot\mathbf{B}\left(\mathbf{r}\right)=0$
and $\boldsymbol{\nabla}\times\mathbf{B}\left(\mathbf{r}\right)=\mathbf{0}$,
so that:

\begin{align}
B_{xx}+B_{zz} & =0,\label{noMonopoles}\\
B_{zx}-B_{xz} & =0.\label{noCurrents}
\end{align}
Hence, we can re-write Eq. (\ref{HSG}) in a form similar to Eq. (\ref{Hint}):

\begin{equation}
H_{\mathrm{SG}}=-\hbar\eta\mathbf{R}\cdot\overleftrightarrow{\mathrm{H}}\cdot\boldsymbol{\sigma},\label{HSG2}
\end{equation}
where the $\eta$ is a small constant with unit inverse time and length
and the constant dimensionless tensor $\overleftrightarrow{\mathrm{H}}$
has the structure:

\[
\overleftrightarrow{\mathrm{H}}\equiv H_{xx}\left(\mathbf{\hat{x}}\mathbf{\hat{x}}-\mathbf{\hat{z}}\mathbf{\hat{z}}\right)+H_{xz}\left(\mathbf{\hat{x}}\mathbf{\hat{z}}+\mathbf{\hat{z}}\mathbf{\hat{x}}\right),
\]
where $H_{xx}$ and $H_{xz}$ are constant scalars.

We will limit the extent of this interaction in two different ways
in the next two subsections: in time and in space. To do this, we
multiply the interaction Hamiltonian from Eq. (\ref{HSG}) by either
$g\left(t\right)$ or $g\left(\mathbf{r}\right)$, functions with
compact support around the origin in time or space. In order not to
change the dimension of the other terms of the Hamiltonian, we will
consider both these functions dimensionless.

The next two subsections approach these two different ways of describing
the limited extent of the measurement interaction. We will first deal
with the simplest case of time dependence, and then compare it with
the more precise case of space dependence, where the collision formalism
from the previous section will be applied.

\subsection{Time-dependent Hamiltonian}

For this first approach, the total Hamiltonian, using Eq. (\ref{HSG2}),
takes the form:

\begin{equation}
H\left(t\right)=H_{0}+H_{\mathrm{SG}}\left(t\right)=\frac{P^{2}}{2m}-\hbar\eta g\left(t\right)\mathbf{R}\cdot\overleftrightarrow{\mathrm{H}}\cdot\boldsymbol{\sigma}.\label{timeDependentH}
\end{equation}

In this case, we can find the final state of the system using a Dyson
series \cite{Dyson}. In this approach, we write the final state of
a system at $t=\tau>0$ in terms of the initial state at $t=-\tau<0$
as an infinite series of iterative integral solutions to the Schrödinger
equation in the interaction picture:

\begin{equation}
\left|\tilde{\Psi}\left(\tau\right)\right\rangle =\sum_{n=0}^{\infty}\left(-\frac{i}{\hbar}\right)^{n}\int_{-\tau}^{\tau}\mathrm{d}t_{1}\int_{-\tau}^{t_{1}}\mathrm{d}t_{2}\ldots\int_{-\tau}^{t_{n-1}}\mathrm{d}t_{n}\tilde{H}\left(t_{1}\right)\tilde{H}\left(t_{2}\right)\ldots\tilde{H}\left(t_{n}\right)\left|\tilde{\Psi}_{\mathrm{i}}\right\rangle .\label{Dyson}
\end{equation}
The series of integrals in Eq. (\ref{Dyson}) is equivalent to the
S-matrix derived in Sec. 3, if we take the limit when $\tau\to\infty$.
This can be seen more clearly by applying $\left\langle \tilde{\Psi}_{\mathrm{f}}\right|$
to the left of Eq. (\ref{Dyson}) and obtaining the transition probability
amplitude $P_{\mathrm{if}}.$

The interaction-picture Hamiltonian $\tilde{H}\left(t\right)$ is:

\[
\tilde{H}\left(t\right)=e^{iH_{0}t/\hbar}H_{\mathrm{SG}}e^{-iH_{0}t/\hbar}=-\hbar\eta g\left(t\right)\left(e^{iP^{2}t/2m\hbar}\mathbf{R}e^{-iP^{2}t/2m\hbar}\right)\cdot\mathcal{H}\cdot\boldsymbol{\sigma}.
\]

As $\mathbf{R}$ and $P^{2}$ commute as $\left[P^{2},\mathbf{R}\right]=-2i\hbar\mathbf{P}$,
and $\left[P^{2},\mathbf{P}\right]=0$, we can write the following
according to the Baker\textendash Campbell-Hausdorff formula:

\begin{equation}
e^{iP^{2}t/2m\hbar}\mathbf{R}e^{-iP^{2}\tau/2m\hbar}=\mathbf{R}+\left(\frac{it}{2m\hbar}\right)\left(-2i\hbar\mathbf{P}\right).\label{BCH}
\end{equation}
Then, according to Eq. (\ref{BCH}), the interaction-picture Hamiltonian
becomes:

\begin{equation}
\tilde{H}\left(t\right)=-\hbar\eta g\left(t\right)\left(\mathbf{R}+\frac{t}{m}\mathbf{P}\right)\cdot\overleftrightarrow{\mathrm{H}}\cdot\boldsymbol{\sigma}.\label{HI with extra term}
\end{equation}

The expansion from Eq. (\ref{Dyson}) can be simplified if we consider
only a weak measurement, that is, for small $\eta$. In this case,
we can keep just the terms of the expansion up to the first order
in $\eta$, arriving at:

\[
\left|\tilde{\Psi}\left(\tau\right)\right\rangle =\left|\tilde{\Psi}_{\mathrm{i}}\right\rangle +i\eta\int_{-\tau}^{\tau}\mathrm{d}t_{1}\:g\left(t_{1}\right)\left(\mathbf{R}+\frac{t_{1}}{m}\mathbf{P}\right)\cdot\overleftrightarrow{\mathrm{H}}\cdot\boldsymbol{\sigma}\left|\tilde{\Psi}_{\mathrm{i}}\right\rangle +O\left(\eta^{2}\right).
\]

If we assume that the function $g\left(t_{1}\right)$ is approximately
symmetric around the origin, the integral of it multiplied by $t_{1}$
from $-\tau$ to $\tau$ vanishes, thus eliminating the term containing
the $\mathbf{P}$ operator. There remains only the term proportional
to $\mathbf{R}$. After a post-selection of a final state of the spin
$\left|\chi_{\mathrm{f}}\right\rangle $, the final relative state
will be, according to the definition in Eq. (\ref{postselection}):

\[
\mathrm{rel}_{\left|\chi_{\mathrm{f}}\right\rangle }\left|\tilde{\Psi}\left(\tau\right)\right\rangle =C\left\langle \chi_{\mathrm{f}}\right.\left|\chi_{\mathrm{i}}\right\rangle \left|\tilde{\Phi}_{\mathrm{i}}\right\rangle +i\eta\int_{-\tau}^{\tau}\mathrm{d}t_{1}\:g\left(t_{1}\right)\mathbf{R}\cdot\mathcal{\overleftrightarrow{\mathrm{H}}}\cdot\left\langle \chi_{\mathrm{i}}\right|\boldsymbol{\sigma}\left|\chi_{\mathrm{f}}\right\rangle \left|\tilde{\Phi}_{\mathrm{i}}\right\rangle +O\left(\eta^{2}\right).
\]
What is left can be approximated by an exponential function that displaces
the system by an amount proportional to the weak vector $\boldsymbol{\sigma}_{\mathrm{w}}$:

\[
\mathrm{rel}_{\left|\chi_{\mathrm{f}}\right\rangle }\left|\tilde{\Psi}\left(\tau\right)\right\rangle =C\left\langle \chi_{\mathrm{f}}\right.\left|\chi_{\mathrm{i}}\right\rangle \exp\left\{ i\eta\,\int_{-\tau}^{\tau}\mathrm{d}t_{1}\:g\left(t_{1}\right)\mathbf{R}\cdot\overleftrightarrow{\mathrm{H}}\cdot\boldsymbol{\sigma}_{\mathrm{w}}\right\} \left|\tilde{\Phi}_{\mathrm{i}}\right\rangle +O\left(\eta^{2}\right).
\]
Hence, we recover the result of the weak vector obtained in Sec. 2.
If $g\left(t\right)$ is unit in the interval of its compact support
of length $T$, its integral will also be $T$, which can be interpreted
as the duration of the interaction with the apparatus:

\begin{equation}
\mathrm{rel}_{\left|\chi_{\mathrm{f}}\right\rangle }\left|\tilde{\Psi}\left(\tau\right)\right\rangle =C\left\langle \chi_{\mathrm{f}}\right.\left|\chi_{\mathrm{i}}\right\rangle \exp\left\{ i\eta T\mathbf{R}\cdot\overleftrightarrow{\mathrm{H}}\cdot\boldsymbol{\sigma}_{\mathrm{w}}\right\} \left|\tilde{\Phi}_{\mathrm{i}}\right\rangle +O\left(\eta^{2}\right).\label{semi-final time-dependent}
\end{equation}

Using the fact that in the interaction picture $\left|\tilde{\Phi}_{\mathrm{i}}\right\rangle =e^{iP^{2}\left(-\tau\right)/2m\hbar}\left|\Phi_{\mathrm{i}}\right\rangle $,
if we substitute the entire wave function from Eq. (\ref{PhiInit})
and use the same procedure from Eq. (\ref{ExpTaylor}), we find the
explicit displacement caused by the weak measurement in the wave function:
\begin{equation}
\mathrm{rel}_{\left|\chi_{\mathrm{f}}\right\rangle }\left|\tilde{\Psi}\left(\tau\right)\right\rangle =C\left\langle \chi_{\mathrm{f}}\right.\left|\chi_{\mathrm{i}}\right\rangle \int_{V_{p}^{\infty}}\mathrm{d}^{3}p\;\varphi_{\mathrm{i}}\left(\mathbf{p}-\hbar\eta T\overleftrightarrow{\mathrm{H}}\cdot\boldsymbol{\sigma}_{\mathrm{w}}\right)e^{-ip^{2}\tau/2m\hbar}\left|\mathbf{p}\right\rangle +O\left(\eta^{2}\right).\label{final time-dependent}
\end{equation}

The result we found at Eq. (\ref{final time-dependent}) is the same
as Eq. (\ref{weak measurement}), which means we reproduced the result
expected of a weak measurement using this perturbative approach to
a Stern-Gerlach experiment that included explicitly the kinetic energy
term in the Hamiltonian. The only difference is the addition of a
phase $e^{-ip^{2}\tau/2m\hbar}$, which does not affect the probability
distribution of the momenta (although it will increase the dispersion
of the wave function in the space of the positions).

Here, the weak vector $\boldsymbol{\sigma}_{\mathrm{w}}$ emerged
from the first term of the S matrix. If we proceeded to higher order
terms, we would find the extant weak tensors from Eq. (\ref{weak tensor}).
In the next subsection, we will re-derive this result using the collision
theory from Sec. 3 and see how the S matrix once again originates
the weak vector. To do this, however, we must use a time-independent
Hamiltonian.

\subsection{Time-independent Hamiltonian}

In this approach, we will consider $g\left(\mathbf{r}\right)$ to
be a function of the space with compact support, so that the Hamiltonian
of the system can be expressed as a time-independent operator:

\begin{equation}
H=H_{0}+\eta H_{\mathrm{int}}=\frac{P^{2}}{2m}-\hbar\eta g\left(\mathbf{R}\right)\mathbf{R}\cdot\mathcal{\overleftrightarrow{\mathrm{H}}}\cdot\boldsymbol{\sigma}.\label{timeIndependentH}
\end{equation}
We will employ this slightly different Hamiltonian to obtain a result
that we will prove to be compatible with the Eq. (\ref{final time-dependent})
that we derived in the previous subsection.

Hence, the probability amplitude given in Eq. (\ref{PifFinal}) becomes:

\begin{equation}
P_{\mathrm{if}}=\left\langle \chi_{\mathrm{f}}\right|\left.\chi_{\mathrm{i}}\right\rangle \left\langle \tilde{\Phi}_{\mathrm{f}}\right|\left.\tilde{\Phi}_{\mathrm{i}}\right\rangle +2\pi i\hbar\eta\left\langle \chi_{\mathrm{f}}\right|\left.\chi_{\mathrm{i}}\right\rangle \left\langle \tilde{\Phi}_{\mathrm{f}}\right|\mathbf{T}\left|\tilde{\Phi}_{\mathrm{i}}\right\rangle \cdot\overleftrightarrow{\mathrm{H}}\cdot\boldsymbol{\sigma}_{\mathrm{w}}+O\left(\eta^{2}\right),\label{PifStep2}
\end{equation}
where the vector operator $\mathbf{T}$ is:

\begin{equation}
\mathbf{T}\equiv\int_{V_{p}^{\infty}}\mathrm{d}^{3}p\int_{V_{p}^{\infty}}\mathrm{d}^{3}p^{\prime}\:\delta\left(\frac{p^{2}}{2m}-\frac{p^{\prime2}}{2m}\right)\left|\mathbf{p}^{\prime}\right\rangle \left\langle \mathbf{p}^{\prime}\right|g\left(\mathbf{R}\right)\mathbf{R}\left|\mathbf{p}\right\rangle \left\langle \mathbf{p}\right|.\label{operator T}
\end{equation}

Some calculus allows us to re-write $\mathbf{T}$ in the following
form:

\[
\mathbf{T}=\frac{1}{2\pi\hbar}\int_{-\infty}^{\infty}\mathrm{d}\tau\int_{V_{p}^{\infty}}\mathrm{d}^{3}p\int_{V_{p}^{\infty}}\mathrm{d}^{3}p^{\prime}\:\tilde{g}\left(\mathbf{p}^{\prime}-\mathbf{p}\right)e^{-iP^{2}\tau/2m\hbar}\left|\mathbf{p}^{\prime}\right\rangle \left\langle \mathbf{p}\right|\mathbf{R}e^{iP^{2}\tau/2m\hbar}.
\]
where $\bar{g}\left(\mathbf{p}\right)$ is the Fourier transform of
$g\left(\mathbf{r}\right)$. The detailed calculations to arrive at
this result can be found in the Appendix B.

Using the Baker-Campbell-Hausdorff identity from Eq. (\ref{BCH}),
we can commute $\mathbf{R}$ and $e^{iP^{2}\tau/2m\hbar}$, arriving
at:

\begin{equation}
\mathbf{T}=\frac{1}{2\pi\hbar}\int_{-\infty}^{\infty}\mathrm{d}\tau\int_{V_{p}^{\infty}}\mathrm{d}^{3}p\int_{V_{p}^{\infty}}\mathrm{d}^{3}p^{\prime}\:\bar{g}\left(\mathbf{p}^{\prime}-\mathbf{p}\right)e^{-iP^{2}\tau/2m\hbar}\left|\mathbf{p}^{\prime}\right\rangle \left\langle \mathbf{p}\right|e^{iP^{2}\tau/2m\hbar}\left(\mathbf{R}-\frac{\tau}{m}\mathbf{P}\right),\label{Tstep4}
\end{equation}

The function $g\left(\mathbf{\mathbf{r}}\right)$ only needs to be
dependent on the coordinate in which the beam of atoms is moving.
Like in the previous subsection, we will call this direction $\mathbf{\hat{y}}$:

\[
g\left(\mathbf{r}\right)=g\left(y\right).
\]

Then, from the Fourier transform given in Eq. (\ref{tildeG}) of thes
Appendix B:

\[
\bar{g}\left(\mathbf{p}\right)=\delta\left(x\right)\delta\left(z\right)\frac{1}{\left(2\pi\hbar\right)}\int_{-\infty}^{\infty}\mathrm{d}y\ e^{-ip_{y}y/\hbar}g\left(y\right)\equiv\delta\left(x\right)\delta\left(z\right)\tilde{g}\left(p_{y}\right).
\]

Using this definition of $\bar{g}\left(p_{y}\right)$, Eq. (\ref{Tstep4})
becomes:

\begin{equation}
\mathbf{T}=\frac{1}{2\pi\hbar}\int_{-\infty}^{\infty}\mathrm{d}\tau\int_{-\infty}^{\infty}\mathrm{d}p_{y}\int_{-\infty}^{\infty}\mathrm{d}p_{y}^{\prime}\:\bar{g}\left(p_{y}^{\prime}-p_{y}\right)e^{-i\left(p_{y}^{\prime2}-p_{y}^{2}\right)\tau/2m\hbar}\left|p_{y}^{\prime}\right\rangle \left\langle p_{y}\right|\left(\mathbf{R}-\frac{\tau}{m}\mathbf{P}\right).\label{Tstep5}
\end{equation}
Here, we can see an extra term containing $\tau\mathbf{P}/m$, which
is the same term we discarded at Eq. (\ref{HI with extra term}) from
the previous subsection, using an argument about the symmetry of $g\left(t\right)$.
In this subsection, we also dismiss this extra term, this time using
an argument involving Dirac's delta:

\begin{equation}
\frac{1}{2\pi\hbar}\int_{-\infty}^{\infty}\mathrm{d}\tau\,e^{i\left(p_{y}^{\prime2}-p_{y}^{2}\right)\tau/2m\hbar}=\delta\left(\frac{p_{y}^{\prime2}}{2m}-\frac{p_{y}^{2}}{2m}\right)=\frac{m}{\left|p_{y}\right|}\delta\left(p_{y}-p_{y}^{\prime}\right)+\frac{m}{\left|p_{y}\right|}\delta\left(p_{y}+p_{y}^{\prime}\right).\label{Dirac energy}
\end{equation}

As we argued in the beginning of this section, the tensor $\mathcal{H}$
has no components in the $\mathbf{\hat{y}}$ direction, so we only
need to take into account components of $\mathbf{T}$ in the $xz$
plane, which contain the components $P_{X}$ and $P_{Z}$ of the operator
$\mathbf{P}$. The typical values of $p_{y}$, the direction of the
beam, will be much larger than the initial values of $p_{x}$ and
$p_{z}$. Therefore, $\mathbf{P}/\left|p_{y}\right|$ when applied
to the initial state $\left|\tilde{\Phi}_{\mathrm{i}}\right\rangle $
will contribute much less than the term proportional to $\mathbf{R}/\left|p_{y}\right|$
\textemdash{} because $\left|p_{x}\right|,\left|p_{z}\right|\ll\left|p_{y}\right|$
\textemdash{} and can thus be neglected.

Hence, replacing Eq. (\ref{Dirac energy}) in Eq. (\ref{Tstep5})
and discarding the extra terms, we find:

\[
\mathbf{T}=\int_{-\infty}^{\infty}\mathrm{d}p_{y}\:\bar{g}\left(0\right)\frac{m}{\left|p_{y}\right|}\left|p_{y}\right\rangle \left\langle p_{y}\right|\mathbf{R}+\int_{-\infty}^{\infty}\mathrm{d}p_{y}\:\bar{g}\left(2p_{y}\right)\frac{m}{\left|p_{y}\right|}\left|-p_{y}\right\rangle \left\langle p_{y}\right|\mathbf{R}.
\]

When $g\left(\mathbf{r}\right)$ is of a much larger scale than the
wave lengths involved in the wave packet $\psi\left(\mathbf{r}\right)$
(which is the case in the Stern-Gerlach apparatus), thus varying very
slowly compared to the frequency of oscillation of the wave packet,
$\bar{g}\left(0\right)$ will typically be much larger than $\bar{g}\left(2p_{y}\right)$.
See, for example, Eq. (\ref{example gpy}) below. For this reason,
we will neglect the second term, keeping only:

\[
\mathbf{T}=\bar{g}\left(0\right)\int_{-\infty}^{\infty}\mathrm{d}p_{y}\frac{m}{\left|p_{y}\right|}\left|p_{y}\right\rangle \left\langle p_{y}\right|\mathbf{R}.
\]

Replacing the extant result in Eq. (\ref{PifStep2}), we find what
is approximately an exponential:

\[
P_{\mathrm{if}}=\left\langle \chi_{\mathrm{f}}\right|\left.\chi_{\mathrm{i}}\right\rangle \left\langle \tilde{\Phi}_{\mathrm{f}}\right|\exp\left\{ 2\pi i\hbar\eta\bar{g}\left(0\right)\left[\int_{-\infty}^{\infty}\mathrm{d}p_{y}\frac{m}{\left|p_{y}\right|}\left|p_{y}\right\rangle \left\langle p_{y}\right|\right]\mathbf{R}\cdot\overleftrightarrow{\mathrm{H}}\cdot\boldsymbol{\sigma}_{\mathrm{w}}\right\} \left|\tilde{\Phi}_{\mathrm{i}}\right\rangle +O\left(\eta^{2}\right).
\]

Consider a compact-support function $g\left(y\right)$ that is a constant
unit value through the length $L$ of its compact support centered
at the origin. In this case, $\bar{g}\left(p_{y}\right)$ is a sinc
function:

\begin{equation}
\bar{g}\left(p_{y}\right)=\frac{1}{2\pi\hbar}\int_{-L/2}^{L/2}\mathrm{d}y\ e^{-ip_{y}y/\hbar}=\frac{1}{\pi}\frac{\sin\left(p_{y}L/2\hbar\right)}{p_{y}},\label{example gpy}
\end{equation}
which in the limit $p_{y}\to0$ becomes $\bar{g}\left(0\right)=L/2\pi\hbar$.
Hence, the probability amplitudes become:

\begin{equation}
P_{\mathrm{if}}=\left\langle \chi_{\mathrm{f}}\right|\left.\chi_{\mathrm{i}}\right\rangle \left\langle \tilde{\Phi}_{\mathrm{f}}\right|\exp\left\{ i\eta L\left[\int_{-\infty}^{\infty}\mathrm{d}p_{y}\frac{m}{\left|p_{y}\right|}\left|p_{y}\right\rangle \left\langle p_{y}\right|\right]\mathbf{R}\cdot\overleftrightarrow{\mathrm{H}}\cdot\boldsymbol{\sigma}_{\mathrm{w}}\right\} \left|\tilde{\Phi}_{\mathrm{i}}\right\rangle +O\left(\eta^{2}\right).\label{final time-independent}
\end{equation}

Let us compare this with the probability amplitude for the time-dependent
case, which we derived in the previous subsection. To do this, we
apply $\left\langle \tilde{\Phi}_{\mathrm{f}}\right|$ to Eq. (\ref{semi-final time-dependent}),
yielding:

\begin{equation}
P_{\mathrm{if}}=C\left\langle \chi_{\mathrm{f}}\right.\left|\chi_{\mathrm{i}}\right\rangle \left\langle \tilde{\Phi}_{\mathrm{f}}\right|\exp\left\{ i\eta T\mathbf{R}\cdot\overleftrightarrow{\mathrm{H}}\cdot\boldsymbol{\sigma}_{\mathrm{w}}\right\} \left|\tilde{\Phi}_{\mathrm{i}}\right\rangle +O\left(\eta^{2}\right).\label{PifT}
\end{equation}

The value $Lm/p_{y}$ that appears in Eq. (\ref{final time-independent})
is the time a particle with momentum $p_{y}$ and mass $m$ takes
to transverse a Stern-Gerlach apparatus of length $L$, and hence
equivalent to the time duration $T$ of the experiment, present in
Eq. (\ref{PifT}). Therefore, both approaches to the weak-measurement
problem from a collision theory perspective yield the same result
as expressed in Eq. (\ref{final time-dependent}). In both cases,
the weak vector originates from the S matrix approximated to the first
order in the perturbation.

\begin{figure}
\includegraphics[width=0.49\textwidth]{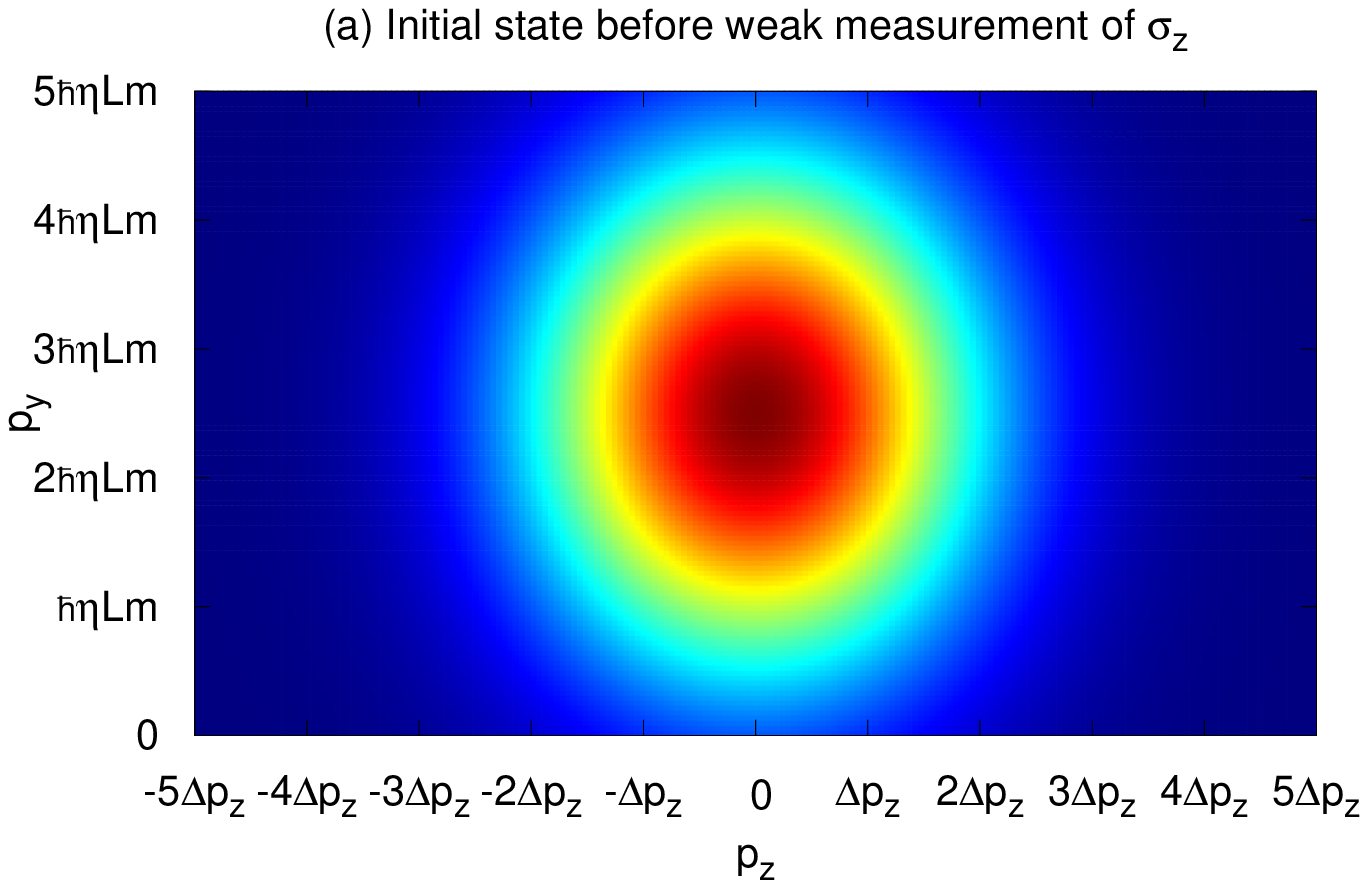}\includegraphics[width=0.49\textwidth]{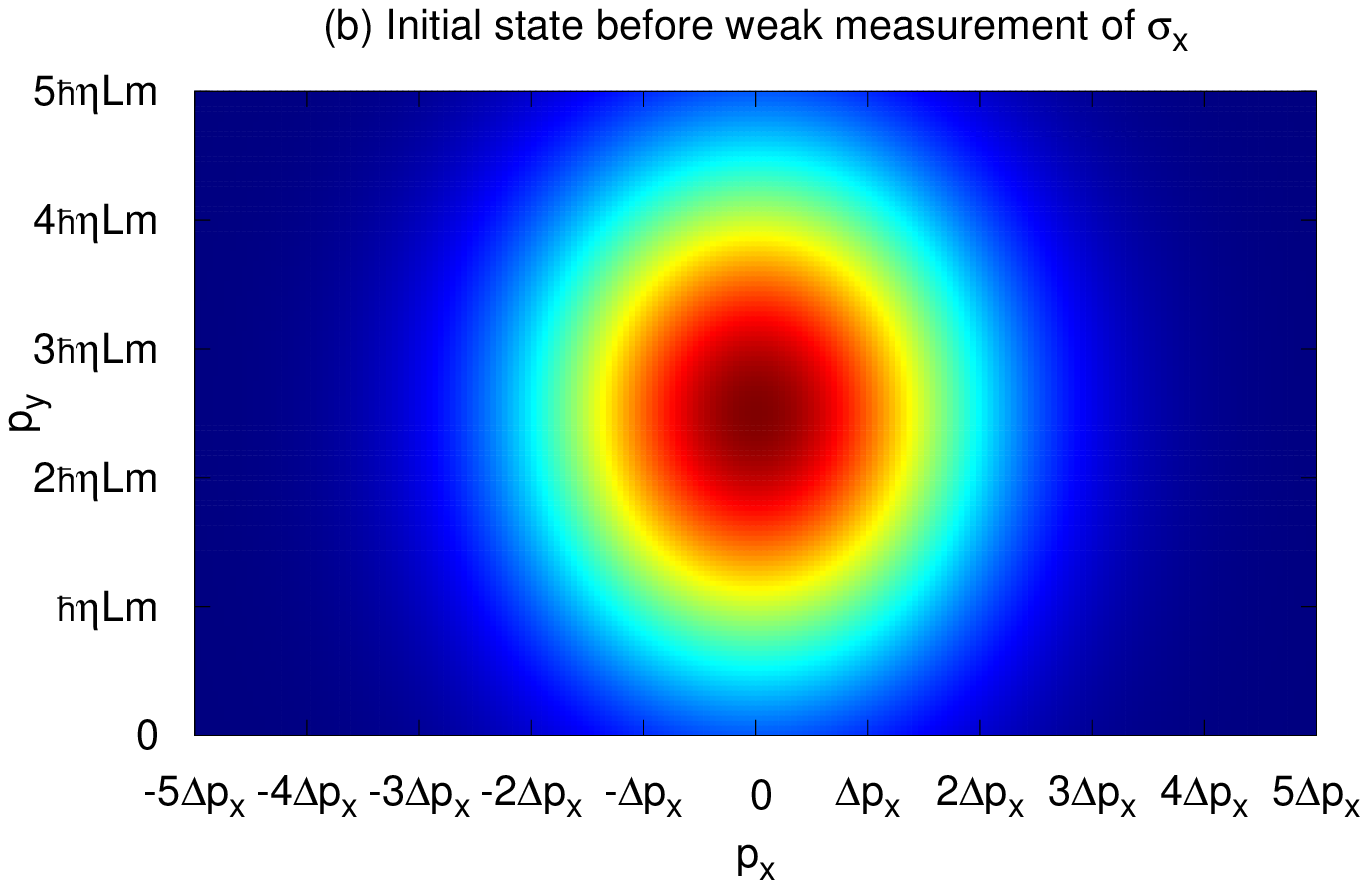}

\includegraphics[width=0.49\textwidth]{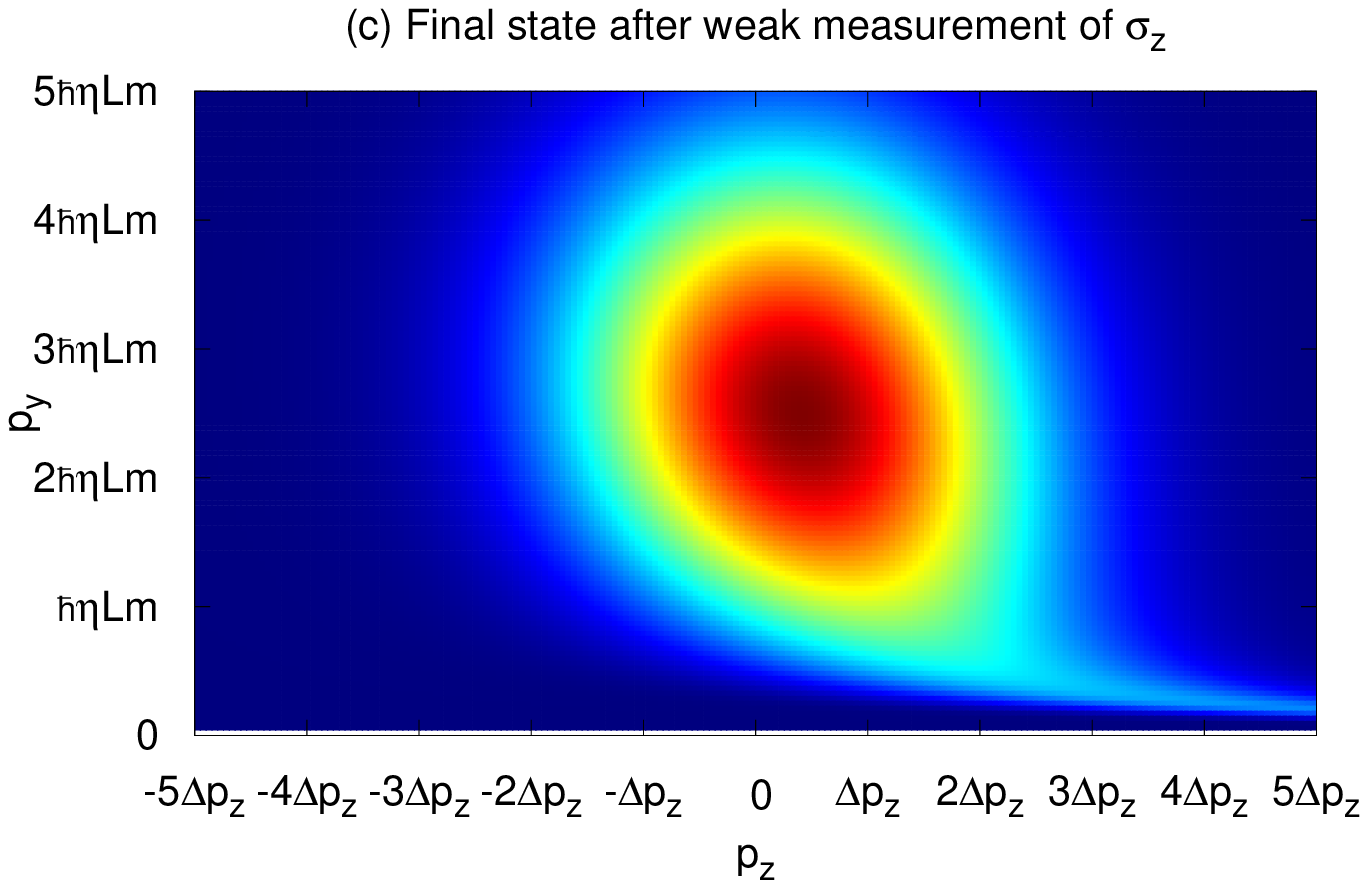}\includegraphics[width=0.49\textwidth]{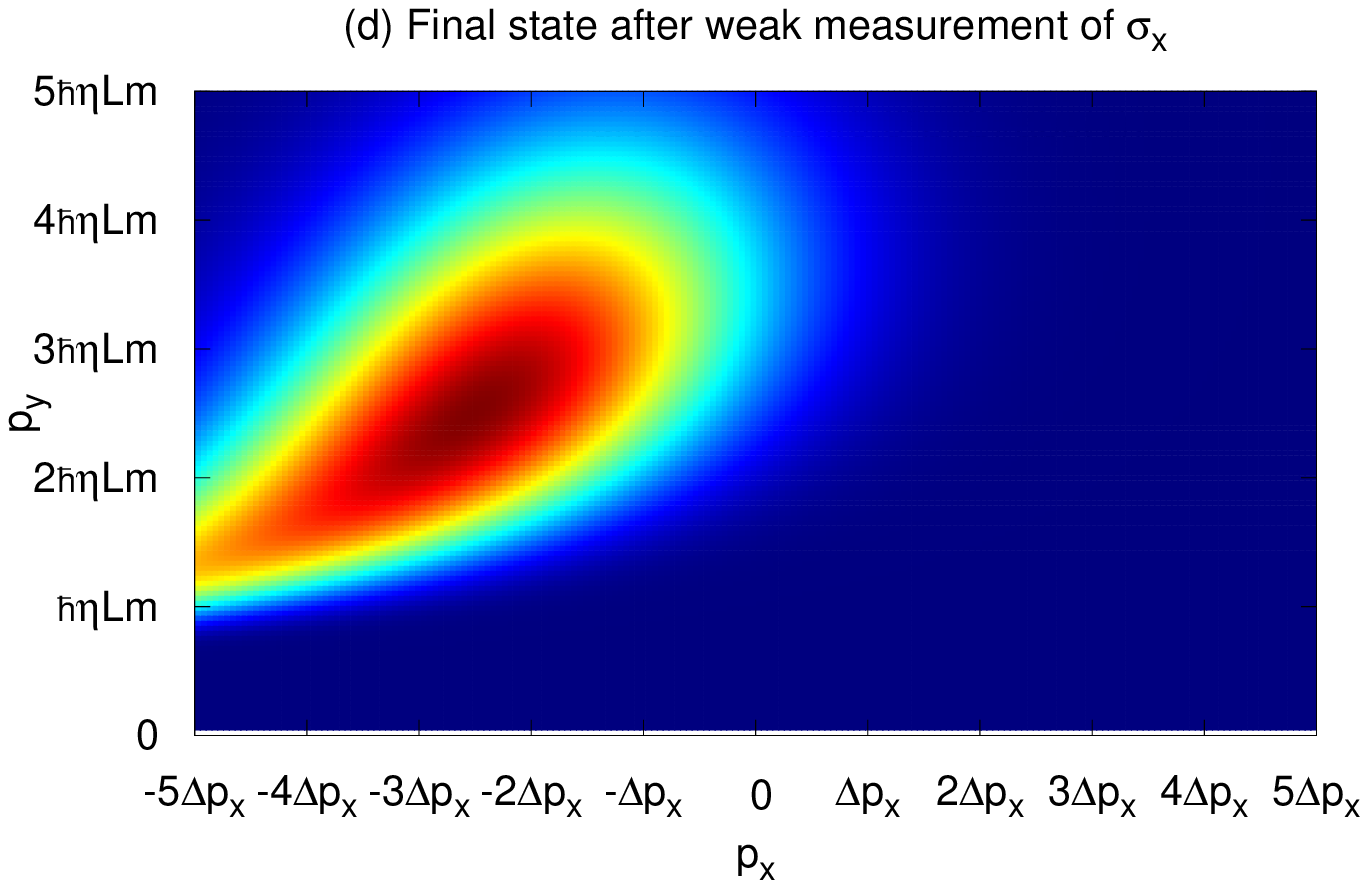}\caption{Heat maps of the Gaussian wave function of the momentum before (a-b)
and after (c-d) a weak measurement of (a-c) $\sigma_{z}$ and (b-d)
$\sigma_{x}$. The same conditions as in Fig. 2 apply, with the additional
assumption that the beam in the $p_{y}$ coordinate is also a Gaussian
centered at $p_{y}=2.5\hbar\eta Lm$. Each section with fixed value
of $p_{y}$ corresponds to an experiment with interaction time $T$
fixed, as found previously in the text. As the velocity of the beam
increases, the particle has less time to interact with the apparatus
and the final wave function remains almost unaltered from its initial
state. As $p_{y}$ goes to zero, however, the interaction becomes
longer and the location of the final Gaussian diverges. }
\end{figure}

The only significant difference between the two treatments is that
the latter includes explicitly the momentum of the beam in the direction
of the movement, a parameter that can be thus controlled to modify
the intensity of the measurement without changing the Stern-Gerlach
apparatus. The way how the momentum $p_{y}$ affects the results of
the weak measurements can be seen in Fig. 3.

\section{Conclusions and perspectives}

In this work we have taken a different approach to the well-known
subject of weak measurements, showing how their \textquotedblleft surprising
results\textquotedblright{} emerge from collision theory. We achieved
our main objective by taking two different paths, a simpler one using
time-dependent Hamiltonians and Dyson series, and another one with
time-independent Hamiltonians and the Lippmann-Schwinger equation.
In the end, we proved that the two results coincide with each other
and with the standard description of weak measurements.

In both cases, the weak values (or weak vectors) were shown to emerge
from the first approximation of the S-matrix. It seems quite clear
from the expansion in Eq. (\ref{Dyson}) that higher-order weak values
(or weak tensors) would come from higher-order terms of the expansion
of the S-matrix, thus demonstrating the intimate connection between
these two concepts \textendash{} one from the theory of weak measurements
and the other from the older collision theory.

During the process of deriving these results, we made fewer approximations
and assumptions than in the ordinary treatment of weak measurements,
and so we have achieved more general results. For one thing, we did
not ignore the effects of the kinetic energy of the beam, which add
an additional factor to Eq. (\ref{HI with extra term}). We found
arguments to ignore this contribution for typical experiments in both
our approaches, but this shows that there may be extreme situations
where effects not usually taken into account may weigh in more heavily.
The path to the exploration of such effects is now open with the help
of collision theory. Likewise for the exploration of the effects of
the dispersion due to $e^{-ip^{2}\tau/2m\hbar}$ in Eq. (\ref{final time-dependent}),
which may affect the form of the wave function in the space of positions.

Moreover, we made approximations to weed out all the terms that contained
higher-order weak values/tensors. A more precise approach using collision
theory may clarify the meaning of these so far neglected variables
and help to estimate their values in different experimental settings.

Finally, in describing the magnetic fields inside the Stern-Gerlach
apparatus in a physically acceptable fashion, we employed
the concept of weak vector, which encapsulates all the information
of possible weak measurements for a spin-$1/2$ particle. It is our
hope that this will prove useful in the better visualization and description
of weak measurements, where in practice all these possible weak values
play a role simultaneously, as we illustrated in Fig. 2(c).

\section*{Acknowledgments}

The authors wish to thank Nícolas André da Costa Morazotti
for his careful reading of the article and comments. C. A. Brasil
acknowledges support from Coordenação de Aperfeiçoamento de Pessoal
de Nível Superior (CAPES) under the Programa Nacional de Pós-Doutorado
(PNPD), and Fundação de Amparo à Pesquisa do Estado de São Paulo (FAPESP),
project number 2011/19848-4, Brazil, and thanks A.
O. Caldeira (IFGW/UNICAMP) and M. H. Y. Moussa (IFSC/USP) for their
hospitality and useful discussions. R. d. J. Napolitano acknowledges
support from Fundação de Amparo à Pesquisa do Estado de São Paulo
(FAPESP), project number 2018/00796-3.

\appendix

\section{Eigenvectors of $H$ and Møller Operators}

In this appendix we will show that the eigenvector of $H$, $\left|\Psi_{\mathbf{\hat{p}},s}^{\pm}\left(E\right)\right\rangle $,
are the result of applying the Møller operators $\Omega^{\pm}$ to
the eigenvector of $H_{0}$, $\left|\Phi_{\hat{\mathbf{p}},s}\left(E\right)\right\rangle $.

We begin by applying $\left(E-H+i\varepsilon\right)^{-1}\left(E-H_{0}+i\varepsilon\right)$
from the left to both sides of Eq. (\ref{LippmannSchwinger}), so
we can write $\left|\Psi_{\mathbf{\hat{p}},s}^{\pm}\left(E\right)\right\rangle $
directly in terms of the eigenvectors of $H_{0}$:

\begin{equation}
\left|\Psi_{\mathbf{\hat{p}},s}^{\pm}\left(E\right)\right\rangle =\lim_{\varepsilon\to0\pm}\frac{1}{E-H+i\varepsilon}i\varepsilon\left|\Phi_{\hat{\mathbf{p}},s}\left(E\right)\right\rangle .\label{LippmannSchwingerClosed}
\end{equation}

The limit above can be written as the following integral:

\[
\lim_{\varepsilon\to0\pm}\frac{1}{E-H+i\varepsilon}i\varepsilon=\lim_{\varepsilon\to0\pm}\frac{\varepsilon}{\hbar}\int_{\mp\infty}^{0}\mathrm{d}t\:\exp\left\{ \left[\frac{\varepsilon}{\hbar}-\frac{i}{\hbar}\left(E-H\right)\right]t\right\} ,
\]
which can be integrated by parts once we recognize that $\left(\varepsilon/\hbar\right)e^{\left(\varepsilon/\hbar\right)t}$
is the time-derivative of $e^{\left(\varepsilon/\hbar\right)t}$:

\begin{align*}
\lim_{\varepsilon\to0\pm}\frac{1}{E-H+i\varepsilon}i\varepsilon & =\lim_{\varepsilon\to0\pm}\left\{ \left.e^{-i\left(E-H\right)t/\hbar}e^{\varepsilon t/\hbar}\right|_{\mp\infty}^{0}-\left(-\frac{i}{\hbar}\right)\left(E-H\right)\int_{\mp\infty}^{0}\mathrm{d}t\;e^{-i\left(E-H\right)t/\hbar}e^{\varepsilon t/\hbar}\right\} \\
 & =1-0-\left(E-H\right)\lim_{\varepsilon\to0\pm}\lim_{t\to\mp\infty}\frac{1-e^{-i\left(E-H\right)t/\hbar}e^{\varepsilon t/\hbar}}{E-H+i\varepsilon},
\end{align*}
allowing us to transform the limit in $\varepsilon$ into a limit
of $t$:

\begin{equation}
\lim_{\varepsilon\to0\pm}\frac{1}{E-H+i\varepsilon}i\varepsilon=\lim_{t\to\mp\infty}e^{-i\left(E-H\right)t/\hbar}.\label{limit identity}
\end{equation}

Using Eq. (\ref{limit identity}) in Eq. (\ref{LippmannSchwingerClosed}),
we can write the eigenvectors of $H$ in terms of the Møller operators:

\[
\left|\Psi_{\mathbf{\hat{p}},s}^{\pm}\left(E\right)\right\rangle =\lim_{t\to\mp\infty}e^{-i\left(E-H\right)t/\hbar}\left|\Phi_{\hat{\mathbf{p}},s}\left(E\right)\right\rangle =\lim_{t\to\mp\infty}\left\{ e^{iHt/\hbar}e^{-iH_{0}t/\hbar}\right\} \left|\Phi_{\hat{\mathbf{p}},s}\left(E\right)\right\rangle =\Omega^{\pm}\left|\Phi_{\hat{\mathbf{p}},s}\left(E\right)\right\rangle .
\]

\section{The Operator $\mathbf{T}$}

In this appendix, we will show how to arrive at the simplified form
of the $\mathbf{T}$ operator presented in Eq. (\ref{Tstep4}).

The matrix element on the second term on the right-hand side of Eq.
(\ref{PifStep2}) becomes, according to the definition Eq. (\ref{operator T}):

\[
\left\langle \tilde{\Phi}_{\mathrm{f}}\right|\mathbf{T}\left|\tilde{\Phi}_{\mathrm{i}}\right\rangle =\frac{1}{2\pi\hbar}\int_{-\infty}^{\infty}\mathrm{d}\tau\int_{V_{p}^{\infty}}\mathrm{d}^{3}p\int_{V_{p}^{\infty}}\mathrm{d}^{3}p^{\prime}\:\left\langle \mathbf{p}\right|\left.\tilde{\Phi}_{\mathrm{i}}\right\rangle \left\langle \tilde{\Phi}_{\mathrm{f}}\right|\left.\mathbf{p}^{\prime}\right\rangle e^{i\left(p^{2}-p^{\prime2}\right)\tau/2m\hbar}\left\langle \mathbf{p}^{\prime}\right|g\left(\mathbf{R}\right)\mathbf{R}\left|\mathbf{p}\right\rangle ,
\]
where we used the integral form of Dirac's delta.

To open the matrix element $\left\langle \mathbf{p}^{\prime}\right|g\left(\mathbf{R}\right)\mathbf{R}\left|\mathbf{p}\right\rangle $,
we can insert a completeness relation for the position eigenvectors
between $g\left(\mathbf{R}\right)$ and $\mathbf{R}$ to project the
former into the function $g\left(\mathbf{r}\right)$ and the latter
into the vector $\mathbf{r}$:

\[
\left\langle \mathbf{p}^{\prime}\right|g\left(\mathbf{R}\right)\mathbf{R}\left|\mathbf{p}\right\rangle =\frac{1}{\left(2\pi\hbar\right)^{3}}\int_{V_{r}^{\infty}}\mathrm{d}^{3}r\ e^{i\left(\mathbf{p}-\mathbf{p}^{\prime}\right)\cdot\mathbf{r}/\hbar}g\left(\mathbf{r}\right)\mathbf{r}.
\]

Notice that $\mathbf{r}e^{i\left(\mathbf{p}-\mathbf{p}^{\prime}\right)\cdot\mathbf{r}/\hbar}=-i\hbar\boldsymbol{\nabla}_{p}e^{i\left(\mathbf{p}-\mathbf{p}^{\prime}\right)\cdot\mathbf{r}/\hbar}$,
so that we can write:

\[
\left\langle \tilde{\Phi}_{\mathrm{f}}\right|\mathbf{T}\left|\tilde{\Phi}_{\mathrm{i}}\right\rangle =-\frac{i}{2\pi}\int_{-\infty}^{\infty}\mathrm{d}\tau\int_{V_{p}^{\infty}}\mathrm{d}^{3}p\int_{V_{p}^{\infty}}\mathrm{d}^{3}p^{\prime}\:\left\langle \mathbf{p}\right|\left.\tilde{\Phi}_{\mathrm{i}}\right\rangle \left\langle \tilde{\Phi}_{\mathrm{f}}\right|\left.\mathbf{p}^{\prime}\right\rangle e^{i\left(p^{2}-p^{\prime2}\right)\tau/2m\hbar}\boldsymbol{\nabla}_{p}\tilde{g}\left(\mathbf{p}^{\prime}-\mathbf{p}\right),
\]
where $\bar{g}\left(\mathbf{p}\right)$ is the Fourier transform of
$g\left(\mathbf{r}\right)$:

\begin{equation}
\bar{g}\left(\mathbf{p}\right)=\frac{1}{\left(2\pi\hbar\right)^{3}}\int_{V_{r}^{\infty}}\mathrm{d}^{3}r\ e^{-i\mathbf{p}\cdot\mathbf{r}/\hbar}g\left(\mathbf{r}\right).\label{tildeG}
\end{equation}

The triple integral in $\mathrm{d}^{3}p$ can be solved using a partial
integration of the gradient:

\begin{align}
-\int_{V_{p}^{\infty}}\mathrm{d}^{3}p\:e^{ip^{2}\tau/2m\hbar}\left\langle \mathbf{p}\right|\left.\tilde{\Phi}_{\mathrm{i}}\right\rangle \boldsymbol{\nabla}_{p}\bar{g}\left(\mathbf{p}^{\prime}-\mathbf{p}\right) & =\int_{V_{p}^{\infty}}\mathrm{d}^{3}p\:\bar{g}\left(\mathbf{p}^{\prime}-\mathbf{p}\right)\boldsymbol{\nabla}_{p}\left[e^{ip^{2}\tau/2m\hbar}\left\langle \mathbf{p}\right|\left.\tilde{\Phi}_{\mathrm{i}}\right\rangle \right]\nonumber \\
 & \quad-\int_{V_{p}^{\infty}}\mathrm{d}^{3}p\:\boldsymbol{\nabla}_{p}\left[e^{ip^{2}\tau/2m\hbar}\left\langle \mathbf{p}\right|\left.\tilde{\Phi}_{\mathrm{i}}\right\rangle \bar{g}\left(\mathbf{p}^{\prime}-\mathbf{p}\right)\right].\label{integral por partes}
\end{align}
The last term on the right-hand side of Eq. (\ref{integral por partes})
can be transformed, via Gauss's theorem, in a surface integral of
$e^{ip^{2}\tau/2m\hbar}\left\langle \mathbf{p}\right|\left.\tilde{\Phi}_{\mathrm{i}}\right\rangle \bar{g}\left(\mathbf{p}^{\prime}-\mathbf{p}\right)$
over the outermost confines of the space of momenta, where both functions
$\tilde{\psi}_{\mathrm{i}}\left(\mathbf{p}\right)$ and $\bar{g}\left(\mathbf{p}\right)$
vanish. Therefore, only the first term of the right-hand side of Eq.
(\ref{integral por partes}) remains:

\begin{align}
\left\langle \tilde{\Phi}_{\mathrm{f}}\right|\mathbf{T}\left|\tilde{\Phi}_{\mathrm{i}}\right\rangle  & =\frac{i}{2\pi}\int_{-\infty}^{\infty}\mathrm{d}\tau\int_{V_{p}^{\infty}}\mathrm{d}^{3}p\int_{V_{p}^{\infty}}\mathrm{d}^{3}p^{\prime}\:\bar{g}\left(\mathbf{p}^{\prime}-\mathbf{p}\right)\left\langle \tilde{\Phi}_{\mathrm{f}}\right|e^{-iP^{2}\tau/2m\hbar}\left|\mathbf{p}^{\prime}\right\rangle \nonumber \\
 & \quad\times\boldsymbol{\nabla}_{p}\left\langle \mathbf{p}\right|e^{iP^{2}\tau/2m\hbar}\left|\tilde{\Phi}_{\mathrm{i}}\right\rangle .\label{Tstep3}
\end{align}

Noticing that

\[
i\hbar\boldsymbol{\nabla}_{p}\left\langle \mathbf{p}\right|e^{iP^{2}\tau/2m\hbar}\left|\tilde{\Phi}_{\mathrm{i}}\right\rangle =\left\langle \mathbf{p}\right|\mathbf{R}e^{iP^{2}\tau/2m\hbar}\left|\tilde{\Phi}_{\mathrm{i}}\right\rangle ,
\]
we can re-write Eq. (\ref{Tstep3}) without $\left|\tilde{\Phi}_{\mathrm{i}}\right\rangle $
and $\left\langle \tilde{\Phi}_{\mathrm{f}}\right|$, if we use the
argument that $\mathbf{T}$ must act on them the same way, no matter
our choice for initial and final states:

\[
\mathbf{T}=\frac{1}{2\pi\hbar}\int_{-\infty}^{\infty}\mathrm{d}\tau\int_{V_{p}^{\infty}}\mathrm{d}^{3}p\int_{V_{p}^{\infty}}\mathrm{d}^{3}p^{\prime}\:\bar{g}\left(\mathbf{p}^{\prime}-\mathbf{p}\right)e^{-iP^{2}\tau/2m\hbar}\left|\mathbf{p}^{\prime}\right\rangle \left\langle \mathbf{p}\right|\mathbf{R}e^{iP^{2}\tau/2m\hbar}.
\]


\begin{thebibliography}{10}
\bibitem{Aharonov1988}Y. Aharonov, D. Z. Albert, L. Vaidman, \emph{How
the result of a measurement of a component of the spin of a spin-1/2
particle can turn out to be 100}, Phys. Rev. Lett. \textbf{60} (1988)
1351.

\bibitem[2]{vN}J. von Neumann, \emph{Mathematical foundations of
quantum mechanics}, Princeton University Press, Princeton, 1955.

\bibitem[3]{ZurekRMP}W. H. Zurek, \emph{Decoherence, einselection,
and the quantum origins of the classical}, Rev. Mod. Phys. \textbf{76}
(2003) 715.

\bibitem[4]{Peres}A. Peres, \emph{Classical interventions in quantum
systems. I. The measuring process}, Phys. Rev. A \textbf{61} (2000)
022116.

\bibitem[5]{Everett}H. Everett, \emph{\textquotedbl{}Relative State\textquotedbl{}
Formulation of Quantum Mechanics,} Rev. Mod. Phys. \textbf{29} (1957)
454.

\bibitem[6]{Wheeler}J. A. Wheeler, \emph{Assessment of Everett's
\textquotedbl{}Relative State\textquotedbl{} Formulation of Quantum
Theory,} Rev. Mod. Phys. \textbf{29} (1957) 463.

\bibitem[7]{Dewitt}B. S. DeWitt, N. Graham, \emph{The many-worlds
interpretation of quantum mechanics}, Princeton University Press,
Princeton, 1973.

\bibitem[8]{JoosBook}E. Joos, H. D. Zeh, C. Kiefer, D. J. W. Giulini,
J. Kupsch, I.-O Stamatescu, \emph{Decoherence and the Appearance of
a Classical World in Quantum Theory}, Springer, Berlin, 2003.

\bibitem[9]{SchlosshauerRMP}M. Schlosshauer, \emph{Decoherence, the
measurement problem, and interpretations of quantum mechanics}, Rev.
Mod. Phys. \textbf{76} (2004) 1267.

\bibitem[10]{SchlosshauerBook}M. Schlosshauer, \emph{Decoherence
and the quantum-to-classical transition}, Springer, Berlin, 2007.

\bibitem[11]{ZurekPT}W. H. Zurek, \emph{Decoherence and the Transition
from Quantum to Classical}, Phys. Today \textbf{44} (1991) 36.

\bibitem[12]{ZurekPTL}W. H. Zurek, \emph{Decoherence and the Transition
from Quantum to Classical\textemdash{} Revisited,} Los Alamos Science
\textbf{27} (2002) 2.

\bibitem[13]{Zeh1}H. D. Zeh, \emph{On the interpretation of measurement
in quantum theory,} Found. Phys. \textbf{1} (1970) 69.

\bibitem[14]{Zeh2}H. D. Zeh, \emph{Toward a quantum theory of observation},
Found. Phys. \textbf{3} (1973) 109.

\bibitem[15]{ZurekPointer}W. H. Zurek, \emph{Pointer basis of quantum
apparatus: Into what mixture does the wave packet collapse?,} Phys.
Rev. D \textbf{24} (1981) 1516.

\bibitem[16]{BrasilEJP}C. A. Brasil, L. A. de Castro, \emph{Understanding
the pointer states,} Eur. J. Phys. \textbf{36} (2015) 065024.

\bibitem[17]{BrasilPRA}C. A. Brasil, L. A. de Castro, R. d. J. Napolitano,
\emph{Protecting a quantum state from environmental noise by an incompatible
finite-time measurement,} Phys. Rev. A \textbf{84} (2011) 022112.

\bibitem[18]{BrasilFP}C. A. Brasil, L. A. de Castro, R. d. J. Napolitano,
\emph{How much time does a measurement take?,} Found. Phys. \textbf{43}
(2013) 642.

\bibitem[19]{BrasilEPJP}C. A. Brasil, L. A. de Castro, R. d. J. Napolitano,
\emph{Efficient finite-time measurements under thermal regimes}, Eur.
Phys. J. Plus \textbf{129} (2014) 206.

\bibitem[20]{Ah1}Y. Aharonov, D. Z. Albert, A. Casher, L. Vaidman,
\emph{Surprising Quantum Effects,} Phys. Lett. A \textbf{124} (1987)
199.

\bibitem[21]{Ah3}Y. Aharonov, L. Vaidman, \emph{Properties of a Quantum
System During the Time Interval Between Two Measurements,} Phys. Rev.
A \textbf{41} (1990) 11.

\bibitem[22]{Duck}I. M. Duck, P. M. Stevenson, E. C. G. Sudarshan,
\emph{The sense in which a \textquotedbl{}weak measurement\textquotedbl{}
of a spin-\textonehalf{} particle's spin component yields a value
100,} Phys. Rev. D \textbf{40} (1989) 2112.

\bibitem[23]{Ah4}Y. Aharonov, P. G. Bermann, J. L. Lebowitz, \emph{Time
Symmetry in the Quantum Process of Measurement}, Phys. Rev. \textbf{134}
(1964) B1410.

\bibitem[24]{Leggett}A. J. Leggett, \emph{Comment on \textquoteleft \textquoteleft How
the result of a measurement of a component of the spin of a spin-1/2
particle can turn out to be 100\textquoteright \textquoteright },
Phys. Rev. Lett. \textbf{62} (1989) 2325.

\bibitem[25]{PeresPRL}A. Peres, \emph{Quantum measurements with postselection},
Phys. Rev. Lett. \textbf{62} (1989) 2326.

\bibitem[26]{Ah5}Y. Aharonov, L. Vaidman, \emph{Aharonov and Vaidman
reply,} Phys. Rev. Lett. \textbf{62} (1989) 2327.

\bibitem[27]{Weinberg}S. Weinberg, \textit{The quantum theory of
fields, volume 1: Foundations}, Cambridge University Press, Cambridge,
1995.

\bibitem[28]{GellMann}M. Gell-Mann, M. L. Goldberger, \emph{The Formal
Theory of Scattering}, Phys. Rev. \textbf{91} (1953) 398.

\bibitem[29]{Merzbacher}E. Merzbacher, \emph{Quantum mechanics},
Wiley, New York, 1997.

\bibitem[30]{Moller}C. Møller, \emph{General Properties of the Characteristic
Matrix in the theory of Elementary Particles I}, Det. K. Danske Vidensk.
Selsk. Mat-fys. Medd. \textbf{23} (1945) 1.

\bibitem[31]{Wheeler2}J. A. Wheeler, \emph{On the Mathematical Description
of Light Nuclei by the Method of Resonating Group Structure}, Phys.
Rev.\textbf{ 52} (1937) 1107.

\bibitem[32]{Solli}D. R. Solli, C. F. McCormick, R. Y. Chiao, S.
Popescu, J. M. Hickmann, \emph{Fast Light, Slow Light, and Phase Singularities:
A Connection to Generalized Weak Values,} Phys. Rev. Lett. \textbf{92}
(2004) 043601.

\bibitem[33]{Williams}R. A. Williams, L. J. LeBlanc, K. Jiménez-García,
M. C. Beeler, A. R. Perry, W. D. Phillips, I. B. Spielman, \emph{Synthetic
Partial Waves in Ultracold Atomic Collisions,} Science \textbf{335}
(2012) 314. 

\bibitem[34]{Hu}W. Hu, \emph{The curious quantum mechanics of pre-
and post-selected ensembles,} Found. Phys. \textbf{20} (1990) 447.

\bibitem[35]{Vaidman}L. Vaidman, \emph{Weak-measurement elements
of reality,} Found. Phys. \textbf{26} (1996) 895.

\bibitem[36]{Kastner}R. E. Kastner, \emph{Weak values and consistent
histories in quantum theory,} St. Hist. Phil. Mod. Phys. \textbf{35}
(2004) 57.

\bibitem[37]{Svensson}B. E. Y. Svensson, \emph{What Is a Quantum-Mechanical
\textquotedblleft Weak Value\textquotedblright{} the Value of?,} Found.
Phys. \textbf{43} (2013) 1193.

\bibitem[38]{Ah6}Y. Aharonov, L. Vaidman, \emph{The Two-State Vector
Formalism: An Updated Review}, Lect. Notes Phys. \textbf{734} (2008)
399.

\bibitem[39]{Ah7}Y. Aharonov, E. Cohen, A. C. Elitzur, \emph{Foundations
and applications of weak quantum measurements,} Phys. Rev. A\textbf{
89} (2014) 052105.

\bibitem[40]{Dressel}J. Dressel, M. Malik, F. M. Miatto, A. N. Jordan,
R. W. Boyd, \emph{Colloquium: Understanding quantum weak values: Basics
and applications,} Rev. Mod. Phys. \textbf{86} (2014) 307.

\bibitem[41]{Ferrie}C. Ferrie, J. Combes, \emph{How the Result of
a Single Coin Toss Can Turn Out to be 100 Heads}, Phys. Rev. Lett.
\textbf{113} (2014) 120404.

\bibitem[42]{Danan}A. Danan, D. Farfurnik, S. Bar-Ad, L. Vaidman,
\emph{Asking Photons Where They Have Been,} Phys. Rev. Lett. \textbf{111}
(2013) 240402.

\bibitem[43]{Acosta}V. M. Acosta, \emph{Quantum information: Strength
of weak measurements}, Nature Physics \textbf{10} (2014) 187.

\bibitem[44]{Blok}M. S. Blok, C. Bonato, M. L. Markham, D. J. Twitchen,
V. V. Dobrovitski, R. Hanson, \emph{Manipulating a qubit through the
backaction of sequential partial measurements and real-time feedback,}
Nature Physics \textbf{10} (2014) 189.

\bibitem[45]{Ibarcq}P. Campagne-Ibarcq, L. Bretheau, E. Flurin, A.
Auffeves, F. Mallet, B. Huard, \emph{Observing Interferences between
Past and Future Quantum States in Resonance Fluorescence,} Phys. Rev.
Lett. \textbf{112} (2014) 180402.

\bibitem[46]{LS1}B. A. Lippmann, J. Schwinger, \emph{Variational
Principles for Scattering Processes. I,} Phys. Rev. \textbf{79} (1950)
469.

\bibitem[47]{LS2}B. A. Lippmann, \emph{Variational Principles for
Scattering Processes. II. Scattering of Slow Neutrons by Para-Hydrogen,}
Phys. Rev. \textbf{79} (1950) 481.

\bibitem[48]{Dyson}F. J. Dyson, \emph{The Radiation Theories of Tomonaga,
Schwinger, and Feynman}, Phys. Rev. \textbf{75} (1949) 486.
\end{thebibliography}
\end{document}